\documentclass[sn-mathphys-num]{sn-jnl}

\usepackage{graphicx}%
\usepackage{multirow}%
\usepackage{amsmath,amssymb,amsfonts}%
\usepackage{amsthm}%
\usepackage{mathrsfs}%
\usepackage[title]{appendix}%
\usepackage{xcolor}%
\usepackage{textcomp}%
\usepackage{manyfoot}%
\usepackage{booktabs}%
\usepackage{algorithm}%
\usepackage{algorithmicx}%
\usepackage{algpseudocode}%
\usepackage{listings}%
\usepackage{lineno,hyperref}
\usepackage{comment}
\newcommand{\comm}[1]{}
\usepackage{color}
\usepackage{subcaption}
\usepackage{cleveref}
\usepackage{float}
\usepackage{newfloat}
\usepackage{arydshln}


\theoremstyle{thmstyleone}%
\theoremstyle{thmstyletwo}%

\theoremstyle{thmstylethree}%

\begin{document}

\title[Article Title]{Are universal machine learning interatomic potentials (uMLIPs) a real alternative to classical interatomic potentials for the allotropes of 2D materials?}


\author*[{~}]{\fnm{Marcin} \sur{Ma\'zdziarz}}\email{mmazdz@ippt.pan.pl}



\affil*[{~}]{\orgname{Institute of Fundamental Technological Research Polish Academy of Sciences}, \orgaddress{\street{Pawi\'nskiego 5B}, \city{Warsaw}, \postcode{02-106}, \country{Poland}}}




\abstract{Specialized machine-learning interatomic potentials (sMLIPs) have not performed well in describing two-dimensional (2D) material allotropes, indicating limited transferability in this domain. Comparisons of interatomic potentials typically focus on calculating the mean absolute error (MAE) against datasets derived from density functional theory (DFT) calculations, rather than assessing the physical correctness of the predictions or whether they satisfy Neumann?s principle. Would universal machine-learning interatomic potentials (uMLIPs) perform better in this context? Furthermore, how does their computational cost on a CPU compare with that of classical interatomic potentials?

}




\maketitle


\section{Introduction}
\label{sec:Int}

In recent years, there has been significant progress in the development of machine learning interatomic potentials (MLIPs) \cite{JACOBS2025101214, Alosious2025, Thiemann2025, Yi2026}.
While the most potentials in the earlier stages were specialized for specific compounds or elements, universal machine learning interatomic potentials (uMLIPs) are gaining popularity because they can be applied to any compound or element. A serious drawback of specialized MLIPs (sMLIPs) was their poor universality/transferability to describe phases of a given compound or element that were not included in the training process, as well as problems with correctly mapping the required symmetry of the tensor physical properties, meaning the failure to satisfy Neumann's principle. For example, it has been found that for some sMLIPs (TurboGAP, GAP20, PANNA) parameterized for carbon \cite{Qamar2023}, the binding energy vs nearest-neighbor distance for certain structures exhibit non-physical oscillatory behavior. For 2D MoS$_2$, the sMLIP (SNAP) did not distinguish between the T and the T' phases \cite{Mazdziarz2021}. For silicene (2D Si) and for stanene (2D Sn), some sMLIPs (qSNAP, SO(3), DP-PBE, DP-SCAN, RANN, MTP) did not reproduce the correct symmetry of certain structures \cite{Mazdziarz2023, Mazdziarz2025}. 
Furthermore, some sMLIPs (ACE, SNAP, POLY, RANN, DP-SCAN, MTP)  failed to reproduce the required isotropy of the stiffness tensor for germanene (2D Ge) and stanene (2D Sn), thereby violating Neumann's principle \cite{Mazdziarz2023b, Mazdziarz2025}.

Universal machine learning interatomic potentials (uMLIPs) are broad-coverage foundation models trained on millions of diverse chemical structures, mostly 3D, to span much of the periodic table. In contrast, specialized MLIPs (sMLIPs) are tightly focused, custom-built models trained exclusively on specific chemistries or targeted applications (like a single alloy or specific surface reaction). 
Various uMLIPs differ in their approach to representing atomic environments \cite{JACOBS2025101214, Alosious2025, Thiemann2025, Yi2026}. Regardless of architectural details they are pre-trained on millions of DFT configurations from datasets like the Materials Project \cite{Jain2013}, Alexandria \cite{Cavignac_2026}, or OMat24 \cite{barroso_omat24}.

It's relatively easy to find publications comparing the performance of different MLIPs. To name a few. The absolute errors in energy and force for five different MLIPs were compared using their own DFT dataset for LiAlO$_2$ in \cite{Roy2026}. For a set of 800 molecules and dimers derived from DFT calculations, the mean absolute error (MAE) of energy differences was calculated for 15 uMLIPs, and their speed and memory requirements on a GPU were estimated in \cite{Eastman2026}. Four uMLIPs were benchmarked \cite{gao2026benchmarkinguniversalmachinelearning} against theoretical data (isotropised bulk modulus, shear modulus, Young's modulus, and Poisson's ratio) for nearly 11,000 elastically stable materials from the Materials Project \cite{Jain2013} database using mean absolute percentage error (MAPE). 
In a subsequent paper \cite{Kazhiyev2026}, based on a dataset comprising 201 organic molecules adsorbed on zinc slabs, four successive uMLIPs were compared in terms of adsorption energy accuracy and fidelity in reproducing the adsorption configuration using the mean absolute error (MAE) and root mean squared error (RMSE) metrics. The authors of the study \cite{Marcial2026} presented a systematic benchmark of nine state-of-the-art uMLIPs and three legacy models through evolutionary crystal structure prediction using RMSE metrics.
Additionally, there are even special platforms dedicated to this. CHIPS-FF (Computational High-Performance Infrastructure for Predictive Simulation-based Force Fields) \cite{Wines2025} allows for the comparison of the mean absolute error for various uMLIPs, calculated with respect to JARVIS-DFT 3D data \cite{Choudhary2020}, for scalar properties such as the lattice constants (a, b and c), volume, bulk modulus, and the C$_{11}$ and C$_{44}$ components of the elasticity tensor, phonon spectra, defect formation energies, surface energies. Related: Matbench Discovery \cite{Riebesell2025}, MOFSimBench \cite{krass2025}, MLIPAudit \cite{wehrhan2025mlipauditbenchmarkingtoolmachine}. 

There are relatively few comprehensive comparisons of the performance of classical potentials with uMLIPs. If such comparisons can be found at all, they are highly selective due to the classical potentials chosen for comparison with uMLIPs \cite{krass2025}. It is not always true that uMLIPs achieve better accuracy than classical potentials. To give a few examples. The PBE-MACE-FT uMLIP model \cite{Fashtali2026} does not offer a more accurate description of the mechanical response of pure graphene than the classical AIREBO potential.
Both the GRACE-2L and MACE-MP0 foundational models \cite{Georgaras2026} perform worse than the classical Stillinger?Weber and Kolmogorov?Crespi models in terms of absolute residual energy per atom for the MoS$_2$/WSe$_2$ bilayer.
The surface energy predictions relative to DFT for 2H- and 3R-MoS$_2$ are worse for MACE-MP-0 and CHGNet uMLIPs \cite{Bilgili2026} than for ReaxFF and Stillinger?Weber potentials.
Six uMLIPs and one classical Universal Force Field (UFF) were compared in terms of adsorption energy errors and interaction energy errors for the modeling of deformation of metal-organic frameworks \cite{Brabson2025}. It was concluded that CHGNet and MACE-MPA-0 outperformed the classical UFF model. The other uMLIP models, namely M3GNet, MACE-MPA-0, and eSEN, performed similarly to or worse than UFF.
For boron clusters and icosahedral boron crystals \cite{Ahmadisharaf2025}, the machine-learning force fields MACE-MP-0, PFP v7.0.0 and SevenNet-MF-OMPA generally outperformed the ReaxFF potentials in terms of both structure and energy for this test set. However, in a few cases, these force fields produced structures with unrealistically short B?B bond lengths and absurdly negative energies. Therefore, although these potentials show great promise, they should be used with more caution than ReaxFF and would probably benefit from refinement using pathological structure.
Traditional potentials generally produce well-shaped, smooth energy landscapes with physically meaningful curvature. This ensures stable force calculations and prevents erratic energy fluctuations that can lead to simulation instabilities. Empirical interatomic potentials encode fundamental physical principles that purely data-driven MLIPs often struggle to learn consistently from finite training data sets \cite{Zheng2026}. 

When comparing the performance of potentials for the properties of materials, the fact that these are often tensor quantities is often overlooked. To illustrate, the elastic constants are not independent entities, but rather components of the stiffness tensor. This imposes mutual constraints on them related to the required symmetry and compliance with the Neumann's Principle. Scalar quantities such as energy and lattice constants are often compared, while for tensor quantities, scalar averages such as the bulk modulus or shear modulus are used.
The UniFFBench framework \cite{Mannan2026} evaluates six state-of-the-art uMLIP models for lattice constants and only five elastic constants, bulk, shear and Young's modulus for the experimental 3D mineral dataset (MinX) using the mean absolute percentage error (MAPE) metric. A similar analysis, but limited only to three elastic constants, was conducted in \cite{lebeda2026benchmarkinguniversalmachinelearninginteratomic} for nine uMLIPs against consistent AIMD reference data for five binary and two high-entropy (HE) 3D carbides. 
The elastic constants and the lattice parameters for $\beta$-Sn obtained from the five first-principles-based settings (DFT/PBE and four PFP uMLIPs modes) and the three MEAM potentials were compared with the experimental reference values \cite{Tatsumi_2026}. Only one uMLIP has a lower MAPE for elastic constants than the best MEAM potential.  

Most interatomic potentials and comparisons are related to three-dimensional (3D) materials, and far fewer involve two-dimensional (2D) materials, especially since uMLIPs exhibit degraded performance when applied to systems of reduced dimensionality compared to 3D bulk compounds \cite{Benedini2025}. These exceptions include Uni2D \cite{wang2026uni2duniversalmachinelearning} within the MatterSim framework, which uses the C2DB database \cite{Haastrup2018}. Unfortunately, here as well, scalar quantities?the effective longitudinal modulus and shear modulus?are used for property prediction.

Classical interatomic potentials rarely yielded non-physical results, such as those that violate the Neumann's principle and generally demonstrated a lower mean absolute percentage error~(MAPE) and significantly better overall performance than specialized machine learning interatomic potentials (sMLIPs). The objective of this study is to examine if the universal machine learning interatomic potentials (uMLIPs) exhibit the same limitations as sMLIPs.

\section{Methodology}
\label{sec:Meth}

All structural and mechanical reference data ($\mathrm{Value^{DFT}}$) for 2D MoS$_2$, Si, Ge, and Sn, i.e., lattice constants, average bond length, average structure height, average cohesive energy, 2D elastic constants, and 2D Kelvin moduli \cite{Mazdziarz2025am} are derived from DFT \cite{DFT-HK, DFT-KS, Hamann2005, Abinit2025} calculations in previous studies \cite{Mazdziarz2021, Mazdziarz2023, Mazdziarz2023b, Mazdziarz2025}.
The same properties determined for the best-performing classical interatomic potentials using molecular statistics (MS) and the  Large-scale Atomic/Molecular Massively Parallel Simulator (LAMMPS) \cite{LAMMPS} are taken from the same studies.
Using a computational methodology analogous to that for classical potentials, for the universal machine learning interatomic potentials (uMLIPs) analyzed and listed in Section \ref{sec:Ip}, a GUI (uMLIP-Interactive) \cite{LEBEDA20266766} for running simulations with uMLIPs ? MACE, CHGNet, Nequix, SevenNet, Orb-v3, MatterSim, UPET, GRACE and the Python atomic simulation environment (ASE) \cite{HjorthLarsen2017} - was used to determine the aforementioned properties and denoted here as ($\mathrm{Value^{potential}}$). The representatives of seven uMLIP families were analyzed. Some families of the UMLIPs include different versions of them. For instance, there are four versions of MACE and two versions of SevenNet. All 11 uMLIPs for MoS$_2$ were analyzed, and a preselection process was conducted to choose the best uMLIP from each family. In the main manuscript, and also for the purpose of providing a overall summary, only one representative uMLIP from each of the seven families was selected. \comm{For Si, Ge, and Sn, as well as for the overall summary, one representative uMLIP from each of the seven families was selected.}
All calculations performed here are in binary64 (double-precision floating-point/FP64) precision. This is because, for standard molecular dynamics trajectories, single precision is sufficient; however, it should be clearly noted that for molecular statics, compilation using double precision is necessary; see \cite{Eastman2010OpenMM, Hess2008}.
In analyzing the structural properties obtained, the Open Visualization Tool~(OVITO) \cite{Stuk2010} was also used. 

A two-step procedure was applied to estimate the computational cost of using the interatomic potentials under study. First, a serial version of LAMMPS (Stable Release from 22 Jul 2025) was compiled with the Intel compiler version 2025.0.0 on a cluster equipped with an Intel Xeon Platinium 8268 processor, 192\,GB of RAM and AlmaLinux. In addition to the classical potentials examined, the CPU version of the universal machine learning interatomic potential MACE was also installed in LAMMPS. The other uMLIPs analyzed either do not have a CPU version for LAMMPS or do not operate properly. 
First, a series of classical molecular dynamics (MD) simulations were performed (several hundred atoms and 1000 time steps, NVE-microcanonical ensemble). Using LAMMPS' built-in function \emph{timesteps/s}, the relative computational cost, normalized to the lowest value, was estimated for a given potential. In the second step, for smaller but analogous tasks to those in the first step, the computational cost of the remaining examined uMLIPs was estimated relative to the MACE potential using the uMLIP-Interactive tool and a laptop. Because MD calculations scale linearly with task size, the relative performance measured per atom and time step is transferable.

\subsection{Performance metrics}
\label{sec:Pm}

When comparing the performance of different interatomic potentials, it is worth saying a few words about the metrics used for this purpose, particularly those recently applied to MLIPs. 
The most common metrics are probably mean absolute error (MAE) and root mean square error (RMSE) \cite{Morrow2023, wehrhan2025mlipauditbenchmarkingtoolmachine, edwards2026benchmarkinguniversalmachinelearnedinteratomic, fu2025learningsmoothexpressiveinteratomic}\comm{, which are calculated for energies, atomic forces, and sometimes also for stresses}:

\begin{equation}
	\mathrm{MAE = \frac{1}{n}\sum_{t=1}^{n}\left |{Value^{DFT}-Value^{potential}}\right|} ,
	\label{eq:mae}
\end{equation}

\begin{equation}
	\mathrm{RMSE = \sqrt{\frac{1}{n}\sum_{t=1}^{n}\left( {Value^{DFT}-Value^{potential}}\right)^2 }} ,
	\label{eq:rmse}
\end{equation}

which are calculated for energies, atomic forces, and sometimes also for stresses.
The typical units of measurement for MAE$_{\mathrm{E,F,S}}$ and RMSE$_{\mathrm{E,F,S}}$ are meV/atom or eV/atom~for energy, meV/\AA or eV/\AA~for forces, and GPa or meV/{\AA}$^3$~for stresses, respectively. When we say that a given error is small, we need to relate it to some value, preferably a physical property. Forces and stresses are not physical properties; they are responses to displacement and strain, respectively. 
In fact, only the energy error measure can be linked to the physical property, i.e., the cohesive energy. However, a problem arises here regarding the reference level because the cohesive energy ranges from $\sim$0.1\,eV/atom for van der Waals (vdW) crystals to $\sim$10\,eV/atom for tungsten (W), a difference of two orders of magnitude. The elastic constants of different materials can differ by two orders of magnitude, for instance, between lithium and diamond \cite{Mazdziarz2024}. This implies that stresses and forces for identical strains can differ by a similar amount.
It has been observed \cite{creed2026openquestionsmachinelearnedinteratomic} that MLIP frameworks perform well within the scope of known chemistry and physics, but struggle when the training data is not representative of the real-world situation. This is largely because they provide an approximate representation of the underlying bonding physics in molecular systems by interpolating within the range of the training data. For rare configurations in typical datasets, standard mean-error objectives such as MAE or RMSE may not accurately reflect practical performance. These losses average over dominant near-equilibrium configurations and can mask catastrophic failures in regions of configuration space that are rare but physically important. Therefore, models with low average errors may still fail to accurately reproduce key atomistic phenomena.

In some cases \cite{Morrow2023, wehrhan2025mlipauditbenchmarkingtoolmachine}, benchmark test results are included in model publications and are used to evaluate physical and chemical properties, usually scalar ones \cite{YuHaochen2024, Focassio2025, Batatia2025, Deng2025}, based on DFT results or experimental data, typically tailored to specific applications. Researchers have also tried to create standard benchmarks to compare models so that they can be used for practical tasks, such as Matbench Discovery \cite{Riebesell2025}, MOFSimBench \cite{krass2025}, MLIPAudit \cite{wehrhan2025mlipauditbenchmarkingtoolmachine}.
Probably the most popular one here, Matbench Discovery, contains 13 separate ML tasks spread across 10 datasets \cite{Dunn2020}.
Unfortunately, these benchmarks are also based on the MAE and RMSE metrics and take into account only scalar physical properties.
To the best of my knowledge, no benchmarks take Neumann's Principle into account, do not verify whether the calculated properties are consistent with the crystal symmetry, and generally do not evaluate the tensor properties of the calculated quantities.

The data thus determined were then taken as reference results and marked as~$\mathrm{Value^{DFT}}$. The same data were then computed by classical molecular statics (MS), as outlined in the Section \ref{sec:Meth}, and the ten interatomic potentials enumerated in {Subsection}~\ref{sec:Ip}. These results were then marked as ~$\mathrm{Value^{potential}}$. Having reference data and those from the MS calculations, the {total} mean absolute percentage error~(MAPE) is specified:

\begin{equation}
	\mathrm{MAPE = \frac{100\%}{n}\sum_{t=1}^{n}\left |\frac{Value^{DFT}-Value^{potential}}{Value^{DFT}}\right|, }
	\label{eq:mape}
\end{equation}
which allows us to quantify the analyzed interatomic potentials.

The use of MAPE as a measure of error requires some comment. The measure is quotient in nature, which excludes reference values close to or equal to zero. For this reason, the average height had to be excluded for the flat structures, since all other reference quantities are significantly different from zero. It is well known that different quantities calculated by DFT have different errors with respect to experimental data, and so, for example, lattice constants are in better agreement with measured ones than elastic moduli or elastic constants. In order to take into account the fact of different accuracy of the calculated quantities, some authors \cite{Jalkanen_2015} propose the introduction of weights in the error measures for these DFT-calculated quantities, the smaller the error the larger the weight. Unfortunately, the choice of these weights is arbitrary and biased. However, it seems reasonable to divide the total mean absolute percentage error~(MAPE) into the structural mean absolute percentage error~(MAPE$^\textrm{S}$), where only structural properties, i.e. lattice parameters, average bond length and average height are considered, and the mechanical mean absolute percentage error~(MAPE$^\textrm{M}$), where only elastic properties and cohesive energy are considered.

To provide an overall estimate of the performance of the potentials under analysis, we also introduce the aggregated total ({MAPE}$_{A}$) as a weighted average for all 19 allotropes (3 MoS$_2$, 5 Si, 4 Ge and 7 Sn) and its 12 properties (4 structural + 8 mechanical) and similarly to before ({MAPE}$^\textrm{S}$$_{A}$) and ({MAPE}$^\textrm{M}$$_{A}$).

\subsection{Interatomic Potentials}
\label{sec:Ip}

The parametrizations of the classical potentials enumerated hereafter were obtained from the~\href{https://www.ctcms.nist.gov/potentials/}{NIST} Interatomic Potentials Repository \cite{NIST} and/or from~LAMMPS code sources and/or through the assistance of the authors of the aforementioned publications. Access to all uMLIPs was made possible by the uMLIP-Interactive tool \cite{LEBEDA20266766}. In LAMMPS on a CPU, the MACE uMLIP uses the interatomic potential style (pair\_style mace) to run \href{https://mace-docs.readthedocs.io/en/latest/index.html}{MACE} machine learning models. 


\begin{enumerate}
	\item \textbf{MACE-MP-0b2 (large)} \cite{Batatia2022mace}: {{MACE} is Higher Order Equivariant Message Passing Neural Networks for Fast and Accurate Force Fields}, using MPTrj Training Set, large variant
	\begin{itemize}
		\item \textbf{MACE-MP-0b2 (small)} \cite{Batatia2022mace}: {{MACE}}, using MPTrj Training Set, small variant
		\item \textbf{MACE-MPA-0 (medium)} \cite{Batatia2022mace}:  {{MACE}}, using MPTrj + sAlex (Subsampled version of the Alexandria dataset) Training Set, medium variant
		\item \textbf{MACE-MATPES-r2SCAN-0 (medium)} \cite{Batatia2022mace}: {{MACE}}, using MATPES-r2SCAN Training Set, medium variant
	\end{itemize}
	\item \textbf{MatterSim-v1.0.0-5M} \cite{yang2024mattersim}: {MatterSim} is models based on the M3GNet architecture, using MatterSim Training Set, large variant
	\item \textbf{CHGNet-0.3.0} \cite{deng2023chgnet}: {CHGNet} is pretrained universal neural network potential for charge-informed atomistic modeling, using MPTrj Training Set, default variant
	\item \textbf{Nequix-MP-1} \cite{koker2025trainingfoundationmodelmaterials}: {Nequix} is a compact E(3)-equivariant potential, using MPTrj Training Set, default variant
	\item \textbf{SevenNet-L3I5} \cite{Park2024}: {SevenNet} is a graph neural network (GNN)-based interatomic potential based on {Nequix}, using MPTrj Training Set, improved variant  
	\begin{itemize}
		\item \textbf{SevenNet-0} \cite{Park2024}: {SevenNet}, using MPTrj Training Set, default variant 
	\end{itemize}
	\item \textbf{ORB-v3 Conservative OMAT} \cite{rhodes2025orbv3atomisticsimulationscale}: {ORB} is a Graph Network Simulator (GNS) augmented with	a smoothed graph attention mechanism, using MPtrj+Alex+OMat24 Training Set \cite{Jain2013,Cavignac_2026,barroso_omat24}, improved variant
	\item \textbf{PET-MAD} \cite{PETMAD2025}: {PET-MAD} is model based on the Point Edge Transformer (PET) architecture, using OMat24 Training Set, default variant 
	
	\noindent\dotfill
	\item \textbf{SW2016-MoS$_2$} \cite{Kandemir_2016}: {SW} is the Stillinger-Weber potential fitted to lattice parameters, distance
	between two chalcogen atoms and elastic constants for SL 1H-MoS$_2$ obtained from DFT calculations
	\item \textbf{REBO-MoS$_2$} \cite{Liang2009,Liang2012}: {REBO} is the reactive many-body potential fitted to structure and energetics of Mo molecules, 3D Mo crystals, 2D Mo structures, small S molecules, and binary Mo-S crystal structures 
	\item \textbf{MEAM2011-Si} \cite{Du2011}: {MEAM} is spline-based modified embedded-atom method potential for Si fitted for silicon interstitials
	\item \textbf{SW1985-Si} \cite{Stillinger1985}: {SW} is the Stillinger-Weber potential fitted to solid and liquid forms of Si
	\item \textbf{SW1986-Ge} \cite{Ding1986}: {SW} is the Stillinger-Weber potential fitted to cohesive energy and elastic constants of {3D} germanium
	\item \textbf{MEAM2018b-Sn} \cite{Ko2018}: {MEAM} is the modified embedded atom method potential whose parameters were optimized based on the force-matching method using the density functional theory (DFT) database of energies and forces of atomic configurations of tin under different conditions 
\end{enumerate}

\section{Results}
\label{sec:Rand}

All the results for the analyzed potentials, uMLIPs, and classical ones for the 19 structures (3 MoS$_2$, 5 Si, 4 Ge and 7 Sn) and its 12 properties (4 structural + 8 mechanical) are all gathered in the \hyperref[sec:SuppI]{Supporting Information}. For MoS$_2$ in Tables~\ref{tab:MoS2_H}-\ref{tab:MoS2_T'}, for silicene in Tables~\ref{tab:Si_F}-\ref{tab:Si_LHD}, for germanene in Tables~\ref{tab:Ge_F}-\ref{tab:Ge_LHD} and for stanene in Tables~\ref{tab:Sn_F}-\ref{tab:Sn_LHD}. The summary results can be found in Table~\ref{tab:MAPE}, and the relative performance measurements for the analyzed potentials in molecular dynamics calculations are presented Table~\ref{tab:time}. For the \textbf{MACE} uMLIP family, the \textbf{MACE MP-0b2 (large)} model had the best performance. For the \textbf{SevenNet} family, it was the \textbf{SevenNet L3I5} model, see Table~\ref{tab:MoS2_T'} for more information. From now on, only these two models will be featured in the main text.  

The first step is to analyze the aggregate results for each of the substances under consideration. The total mean absolute percentage error $\sum$MAPE$_\textrm{3MoS$_2$}$ for the three primary polymorphs of two-dimensional molybdenum disulfide (2D MoS$_2$) was obtained by the classical \textbf{REBO} (104.840) potential, slightly ahead of \textbf{MACE} uMLIP (106.435). The \textbf{ORB} uMLIP (1.746) best replicated pure structural properties, while the \textbf{MACE} (183.3) best replicated pure mechanical properties, see Table~\ref{tab:MMoS2_Tp}.
The total mean absolute percentage error $\sum$MAPE$_\textrm{5Si}$ for the five polymorphs of silicene (2D Si) was obtained by the classical \textbf{MEAM} (82.553) potential, slightly ahead of \textbf{MACE} uMLIP (84.601). The \textbf{ORB} uMLIP (13.58) best replicated pure structural properties, while the \textbf{MACE} (157.409) best replicated pure mechanical properties, see Table~\ref{tab:MSi_LHD}.
The total mean absolute percentage error $\sum$MAPE$_\textrm{4Ge}$ for the four polymorphs of germanene (2D Ge) was obtained by the \textbf{SevenNet} uMLIP (46.607) potential. The \textbf{ORB} uMLIP (11.469) best replicated pure structural properties, while the \textbf{SevenNet} (96.332) best replicated pure mechanical properties, see Table~\ref{tab:MGe_LHD}.
The total mean absolute percentage error $\sum$MAPE$_\textrm{7Sn}$ for the seven polymorphs of stanene (2D Sn) was obtained by the \textbf{MatterSim} uMLIP (200.270) potential. The \textbf{ORB} uMLIP (12.919) best replicated pure structural properties, while the \textbf{Nequix} (235.242) best replicated pure mechanical properties, see Table~\ref{tab:MSn_LHD}.
The aggregated {total} mean absolute percentage error $\sum${MAPE}$_\textrm{A}$ as a weighted average for all 19 allotropes (3 MoS$_2$, 5 Si, 4 Ge and 7 Sn) was obtained by the \textbf{Nequix} uMLIP (117.269) potential. The \textbf{ORB} uMLIP (11.024) best replicated pure structural properties, while the \textbf{MACE} (203.545) best replicated pure mechanical properties, see Table~\ref{tab:MMAPE}. As you can see, modeling mechanical properties is much more difficult than modeling structural properties. Since the stiffness tensor is the second derivative of energy with respect to strain, using the Hessian when optimizing the potential \cite{Rodriguez2025} might improve its performance with regard to mechanical properties. 

The error measures provided are only a rough estimate of the quality of the analyzed potentials; they do not reflect the physical correctness of the results obtained. Averages obscure the details.
For 2D MoS$_2$, the \textbf{ORB}, \textbf{PET-MAD} and \textbf{Nequix} uMLIPs generally could not reproduce the required isotropy of the elasticity tensor correctly, see Tables~\ref{tab:MoS2_H}-\ref{tab:MoS2_T}. 
For silicene, the \textbf{CHGNet}, \textbf{ORB}, \textbf{PET-MAD} and \textbf{Nequix} had the same problems with satisfying Neumann's principle, see Tables~\ref{tab:Si_F}-\ref{tab:Si_LHD} and \textbf{Nequix} uMLIP reproduced a flat structure, not a proper low-buckled one, see Table~\ref{tab:Si_LB}. 
For germanene, the \textbf{CHGNet} and \textbf{PET-MAD} had the same problems with the required isotropy of the elasticity tensor, see Tables~\ref{tab:Ge_F}-\ref{tab:Ge_LHD}.
For stanene, the \textbf{CHGNet}, \textbf{SevenNet}, \textbf{ORB} and \textbf{PET-MAD} had the same problems with the required isotropy of the elasticity tensor, see Tables~\ref{tab:Sn_F}-\ref{tab:Sn_LHD}. Additionally \textbf{CHGNet} and \textbf{ORB} uMLIP did not reproduce the correct symmetry of the low-buckled (LB) polymorph (a$\neq$b), see Table~\ref{tab:Sn_LB}. The \textbf{CHGNet} uMLIP incorrectly indicates a lack of mechanical stability of high-buckled (HB) phase, see Table~\ref{tab:Sn_HB}. 
Only two uMLIPs, namely \textbf{MACE} and \textbf{MatterSim}, never showed non-physical results for all 19 structures, and they are more recommendable than \textbf{Nequix}, even though \textbf{Nequix} had a lower aggregated total mean absolute percentage error.
If we look at the average performance of the best-selected classical potentials versus uMLIPs (Table~\ref{tab:MMAPE}), we do not see any particular advantage of the latter under the methodology used. Most uMLIPs perform worse than classical potentials and often have trouble satisfying Neumann's principle, meaning they are unable to produce physically correct results. For the specific materials discussed here, namely MoS$_2$ and 2D Si, the classical potentials, specifically \textbf{REBO} and \textbf{MEAM}, proved to be the best.
Classical potentials have not been optimized for a large number of structures or features; most often, they are limited to a single phase. However, unlike uMLIPs, they have physics encoded within them, and this is the reason for their resistance to non-physical predictions.  
Since the performance of potential analysis takes into account 12 properties (4 structural and 8 mechanical), we can also estimate the expected average relative error per property. We can expect an error ranging from 3 to 14 \% for the analyzed interatomic potentials for structural properties and from 25 to 75 \% for mechanical properties.

\begin{table}[htbp]
	\caption{Structural and mechanical properties of SL 1T'-MoS$_2$ from molecular calculations: lattice parameters a, b\,(\AA), \comm{and $\gamma$\,($^\circ$)} average cohesive energy $E_{c}$\,(eV/atom), average bond length $d$\,(\AA), average height $h$\,(\AA), 2D elastic constants $\mathrm{C_{ij}}$\,(N/m), 2D Kelvin moduli $K_i$\,(N/m), {total} mean absolute percentage error~(MAPE)\,(\%), {structural mean absolute percentage error~(MAPE$^\textrm{S}$)\,(\%), mechanical mean absolute percentage error~(MAPE$^\textrm{M}$)\,(\%)}.}
	\label{tab:MMoS2_Tp}
	\centering
	\setlength{\tabcolsep}{1.0pt} 
		\begin{tabular}{c c c c c c c c c c}
			\hline
			\textbf{Model} &\textbf{DFT}& \textbf{REBO} & \textbf{MACE} & \textbf{MatterSim} & \textbf{CHGNet} & \textbf{Nequix} & \textbf{SevenNet} &\textbf{ORB} &\textbf{PET-MAD}\\
			\hline      a   & 5.751 & 5.563 & 5.939 & 5.870 & 5.763 & 5.918 & 5.686 & 5.755 & 5.799\\
			b   & 3.177 & 3.245 & 3.222 & 3.149 & 3.215 & 3.178 & 3.212 & 3.172 & 3.143\\
			$-E_{c}$        & 5.562 & 6.934 & 4.887 & 4.779 & 5.119 & 4.807 & 4.977 & 3.798 & 6.585\\
			$d$             & 2.415 & 2.468 & 2.456 & 2.435 & 2.429 & 2.448 & 2.459 & 2.438 & 2.423\\
			$h$             & 3.364 & 3.781 & 3.330 & 3.378 & 3.436 & 3.269 & 3.436 & 3.457 & 3.375\\
			$\mathrm{C_{11}}$ & 68.1 & 56.8 & 105.0 & 130.2 & 125.6 & 117.9 & 134.2 & 147.2 & 153.5\\
			$\mathrm{C_{22}}$ & 78.9 & 113.0& 92.0  & 122.2 & 126.2 & 105.3 & 118.1 & 121.1 & 170.0\\
			$\mathrm{C_{12}}$ & 18.2 & 23.1 & 49.6  &  45.8 &  55.6 &  45.1 &  55.8 &  53.5 &  52.5\\
			$\mathrm{C_{33}}$ & 43.2 & 70.5 & 32.8  &  45.5 &  35.8 &  33.3 &  42.6 &  56.7 &  32.7\\
			$K_{I}$           & 90.9 &121.3 & 147.7 & 171.8 & 181.6 & 156.3 & 181.3 & 186.0 & 213.6\\
			$K_{II}$          & 56.1 & 48.5 &  49.3 &  80.5 &  70.3 &  66.9 &  71.0 &  82.2 & 109.8\\
			$K_{III}$         & 86.4 &141.0 &  65.5 &  91.1 &  71.5 &  66.6 &  85.1 & 113.4 &  65.3\\
			MAPE$_\textrm{T'}$              & & 25.395 & 32.169 & 38.280 & 43.470 & 34.342 & 41.573 & 51.069 & 60.837\\
			{MAPE$^\textrm{S}_\textrm{T'}$} & & 5.000  & 1.840  &  1.040 &  1.033 &  1.783 &  1.557 &  0.984 &  0.646\\
			{MAPE$^\textrm{M}_\textrm{T'}$} & & 54.073 & 56.414 & 67.439 & 70.639 &	60.785 & 69.451 & 99.883 & 104.756\\
			\hline
			$\sum$MAPE$_\textrm{3MoS$_2$}$             & & 104.840 & 106.435 & 144.024 & 183.360 & 120.122 & 153.305 & 145.986 & 181.081\\
			{$\sum$MAPE$^\textrm{S}_\textrm{3MoS$_2$}$}& &   7.736 &   3.549 &   2.490 &   2.598 &   2.996 &   2.959 &   1.746 &   3.246\\
			{$\sum$MAPE$^\textrm{M}_\textrm{3MoS$_2$}$}& & 213.100 & 183.300 & 244.662 & 290.530 & 207.725 & 251.154 & 288.422 & 312.532\\
			\hline
		\end{tabular}
	\comm{\begin{tablenotes}[para,flushleft]
			\begin{tablenotes}[para]
				\small
		\end{tablenotes} }
	\end{table}

\begin{table}[htbp]
	\caption{Structural and mechanical properties of large honeycomb dumbbell (LHD) silicene from molecular calculations: lattice parameters a, b\,(\AA), \comm{and $\gamma$\,($^\circ$)} average cohesive energy $E_{c}$\,(eV/atom), average bond length $d$\,(\AA), average height $h$\,(\AA), 2D elastic constants $\mathrm{C_{ij}}$\,(N/m), 2D Kelvin moduli $K_i$\,(N/m), {total} mean absolute percentage error~(MAPE)\,(\%), {structural mean absolute percentage error~(MAPE$^\textrm{S}$)\,(\%), mechanical mean absolute percentage error~(MAPE$^\textrm{M}$)\,(\%)}.}
	\label{tab:MSi_LHD}
	\centering
	\setlength{\tabcolsep}{1.0pt} 
		\begin{tabular}{c c c c c c c c c c}
			\hline
			\textbf{Model} &\textbf{DFT}& \textbf{MEAM} & \textbf{MACE} & \textbf{MatterSim} & \textbf{CHGNet} & \textbf{Nequix} & \textbf{SevenNet} &\textbf{ORB} &\textbf{PET-MAD}\\
			\hline      a   & 7.334 & 7.363 & 7.474 & 7.408 & 7.556\textsuperscript{$\dag$} & 7.541\textsuperscript{$\dag$} & 7.542 & 7.461 & 7.547\\
			b   & 7.334 & 7.363 & 7.474 & 7.408 & 7.555\textsuperscript{$\dag$} & 7.540\textsuperscript{$\dag$} & 7.542 & 7.461 & 7.547\\
			$-E_{c}$        & 4.769 & 3.911 & 4.180 & 3.429 & 3.737 & 3.784 & 3.634 & 3.241 & 4.531\\
			$d$             & 2.357 & 2.407 & 2.415 & 2.542 & 2.399 & 2.537 & 2.411 & 2.499 & 2.379\\
			$h$             & 2.683 & 2.857	& 2.678 & 2.825 & 2.578 & 2.800 & 2.652 & 2.656 & 2.457\\
			$\mathrm{C_{11}}$ & 104.5 & 88.0  & 72.1 & 72.6 & 65.9\textsuperscript{$\ddag$} & 72.3 & 83.2 & 84.2 & 80.2\textsuperscript{$\ddag$}\\
			$\mathrm{C_{22}}$ & 104.5 & 88.0  & 72.1 & 72.6 & 66.1\textsuperscript{$\ddag$} & 72.3 & 83.2 & 84.2 & 80.9\textsuperscript{$\ddag$}\\
			$\mathrm{C_{12}}$ & 52.7  & 29.9  & 17.7 & 22.7 & 18.0 & 19.8 & 21.7 & 24.1 & 25.1\\
			$\mathrm{C_{33}}$ & 25.9  & 29.0  & 27.2 & 25.0 & 24.2 & 26.2 & 30.8 & 30.1 & 27.6\\
			$K_{I}$           & 157.2 & 117.9 & 89.7 & 95.3 & 84.0 & 92.0 &104.9 &108.4 &105.6\\
			$K_{II}$          & 51.8  & 58.1  & 54.4 & 49.9 & 48.0\textsuperscript{$\ddag$} & 52.5 & 61.5\textsuperscript{$\ddag$} & 60.1 & 55.5\textsuperscript{$\ddag$}\\
			$K_{III}$         & 51.8  & 58.1  & 54.4 & 49.9 & 48.5\textsuperscript{$\ddag$} & 52.5 &	61.6\textsuperscript{$\ddag$} & 60.1 & 55.2\textsuperscript{$\ddag$}\\
			MAPE$_\textrm{LHD}$              & & 13.636 & 17.113 & 17.615 & 19.969 & 17.329 & 18.519 & 17.907 & 14.286\\
			{MAPE$^\textrm{S}_\textrm{LHD}$} & & 2.351  &  1.624 &  3.787 &  2.945 &  4.411 &  2.282 &  2.631 &  3.791\\
			{MAPE$^\textrm{M}_\textrm{LHD}$} & & 32.762 & 34.109 & 45.599 & 44.715 & 39.269 & 44.483 & 49.571 & 23.275\\
			\hline
			$\sum$MAPE$_\textrm{5Si}$             & & 82.553 & 84.601 & 89.123 & 215.614 & 130.171 & 113.883 & 118.370 & 142.379\\
			{$\sum$MAPE$^\textrm{S}_\textrm{5Si}$}& & 22.811 & 35.319 & 17.878 & 144.683 &  38.480 &  21.035 &  13.580 &  36.816\\
			{$\sum$MAPE$^\textrm{M}_\textrm{5Si}$}& &174.224 & 157.409&229.224 & 330.122 & 249.477 & 250.696 & 292.434 & 214.853\\
			\hline
		\end{tabular}
	\begin{tablenotes}[para]
		\small
		\textsuperscript{$\dag$} Potential does not reproduce the correct symmetry of the structure (a$\neq$b), \\ 
		\textsuperscript{$\ddag$} {Potential does not reproduce the required isotropy of the elasticity tensor ($\mathrm{C_{11}}\neq\mathrm{C_{22}}$}~\&$/$$\wedge$~$K_{II}$$\neq$$K_{III}$). 
	\end{tablenotes}
\end{table}

\begin{table}[htbp]
	\caption{Structural and mechanical properties of large honeycomb dumbbell (LHD) germanene from molecular calculations: lattice parameters a, b\,(\AA), \comm{and $\gamma$\,($^\circ$)} average cohesive energy $E_{c}$\,(eV/atom), average bond length $d$\,(\AA), average height $h$\,(\AA), 2D elastic constants $\mathrm{C_{ij}}$\,(N/m), 2D Kelvin moduli $K_i$\,(N/m), {total} mean absolute percentage error~(MAPE)\,(\%), {structural mean absolute percentage error~(MAPE$^\textrm{S}$)\,(\%), mechanical mean absolute percentage error~(MAPE$^\textrm{M}$)\,(\%)}.}
	\label{tab:MGe_LHD}
	\centering
	\setlength{\tabcolsep}{1.0pt} 
		\begin{tabular}{c c c c c c c c c c}
			\hline
			\textbf{Model} &\textbf{DFT}& \comm{\textbf{SW1986}} \textbf{SW} & \textbf{MACE} & \textbf{MatterSim} & \textbf{CHGNet} & \textbf{Nequix} & \textbf{SevenNet} &\textbf{ORB} &\textbf{PET-MAD}\\
			\hline      a   & 7.667 & 7.791 & 7.800 & 7.874 & 7.828 & 7.914 & 7.915 & 7.885 & 7.849\\
			b   & 7.667 & 7.791 & 7.800 & 7.874 & 7.828 & 7.914 & 7.915 & 7.885 & 7.849\\
			$-E_{c}$        & 3.863 & 2.844 & 3.448 & 2.859 & 3.214 & 3.246 & 3.345 & 2.501 & 4.329\\
			$d$             & 2.485 & 2.642 & 2.589 & 2.569 & 2.542 & 2.537 & 2.572 & 2.542 & 2.498\\
			$h$             & 2.875 & 3.449 & 3.209 & 3.025 & 2.950 & 2.800 & 2.981 & 2.880 & 2.715\\
			$\mathrm{C_{11}}$ & 57.7& 76.3  & 39.4  & 47.0  &  42.2\textsuperscript{$\ddag$} &  53.9 &  43.0 & 42.7  & 58.3\textsuperscript{$\ddag$}\\
			$\mathrm{C_{22}}$ & 57.7& 76.3  & 39.4  & 47.0  &  42.3\textsuperscript{$\ddag$} &  53.9 &  43.0 & 42.7  & 58.6\textsuperscript{$\ddag$}\\
			$\mathrm{C_{12}}$ & 20.7& 35.4  & 12.5  & 12.2  &  11.0 &  17.8 &  16.4 & 11.2  & 14.7\\
			$\mathrm{C_{33}}$ & 18.5& 20.4  & 13.4  & 17.4  &  15.7 &  18.1 &  13.3 & 15.8  & 21.9\\
			$K_{I}$           & 78.4& 111.7 & 51.9  & 59.2  &  53.3 &  71.7 &  59.5 & 53.9  & 73.2\\
			$K_{II}$          & 37.0& 40.9  & 26.9  & 34.8  &  31.2\textsuperscript{$\ddag$} &  36.2 &  26.6 & 31.5\textsuperscript{$\ddag$}  & 43.8\textsuperscript{$\ddag$}\\
			$K_{III}$         & 37.0& 40.9  & 26.9  & 34.8  &  31.4\textsuperscript{$\ddag$} &  36.2 &  26.6 & 31.7\textsuperscript{$\ddag$}  & 43.7\textsuperscript{$\ddag$}\\
			MAPE$_\textrm{LHD}$              & & 22.085 & 20.727 & 13.416 & 17.027 & 5.845 & 17.282 & 18.031 & 9.634\\
			{MAPE$^\textrm{S}_\textrm{LHD}$} & & 7.378  &  4.811 &  3.502 &  2.273 & 2.803 &  3.412 &  2.044 & 2.714\\
			{MAPE$^\textrm{M}_\textrm{LHD}$} & & 49.210 & 36.731 & 37.869 & 37.008 &19.342 & 34.273 & 52.467 &22.145\\
			\hline
			$\sum$MAPE$_\textrm{4Ge}$             & &84.125  & 63.451 & 61.761 & 193.991 & 43.854 & 46.607 & 73.206 & 93.547\\
			{$\sum$MAPE$^\textrm{S}_\textrm{4Ge}$}& & 28.475 & 14.510 & 18.551 &  44.555 & 37.806 & 21.595 & 11.469 & 17.346\\
			{$\sum$MAPE$^\textrm{M}_\textrm{4Ge}$}& &182.030 & 117.354&159.380 & 307.893 & 94.205 & 96.332 &210.261 &163.289\\
			\hline
		\end{tabular}
	\begin{tablenotes}[para]
		\small
		\textsuperscript{$\ddag$} {Potential does not reproduce the required isotropy of the elasticity tensor ($\mathrm{C_{11}}\neq\mathrm{C_{22}}$}~\&$/$$\wedge$~$K_{II}$$\neq$$K_{III}$). 
	\end{tablenotes}
\end{table}

\begin{table}[htbp]
	\caption{Structural and mechanical properties of large honeycomb dumbbell (LHD) stanene from molecular calculations: lattice parameters a, b\,(\AA), \comm{and $\gamma$\,($^\circ$)} average cohesive energy $E_{c}$\,(eV/atom), average bond length $d$\,(\AA), average height $h$\,(\AA), 2D elastic constants $\mathrm{C_{ij}}$\,(N/m), 2D Kelvin moduli $K_i$\,(N/m), {total} mean absolute percentage error~(MAPE)\,(\%), {structural mean absolute percentage error~(MAPE$^\textrm{S}$)\,(\%), mechanical mean absolute percentage error~(MAPE$^\textrm{M}$)\,(\%)}.}
	\label{tab:MSn_LHD}
	\centering
	\setlength{\tabcolsep}{1.0pt} 
		\begin{tabular}{c c c c c c c c c c}
			\hline
			\textbf{Model} &\textbf{DFT}& \textbf{MEAM}  & \textbf{MACE} & \textbf{MatterSim} & \textbf{CHGNet} & \textbf{Nequix} & \textbf{SevenNet} &\textbf{ORB} &\textbf{PET-MAD}\\
			\hline      a   & 8.890	& 8.907 & 9.244 & 9.286 & 9.206 & 9.264 & 9.210 & 9.107 & 8.962\\
			b   & 8.890	& 8.907 & 9.244 & 9.286 & 9.206 & 9.264 & 9.210 & 9.107 & 8.962\\
			$-E_{c}$        & 3.062 &  2.830 & 2.903 & 2.250 & 2.331 & 2.787 & 2.803 & 1.989 & 3.063\\
			$d$             & 2.897 &  2.894 & 2.959 & 2.965 & 2.942 & 2.960 & 2.943 & 2.937 & 2.907\\
			$h$             & 3.404 &  3.617 & 3.277 & 3.274 & 3.270 & 3.261 & 3.241 & 3.348 & 3.273\\
			$\mathrm{C_{11}}$ & 33.0&  16.7  & 29.3  &  30.9 & 29.1  & 30.0  & 28.5  & 40.5  & 80.4\textsuperscript{$\ddag$}\\
			$\mathrm{C_{22}}$ & 33.0&  16.7  & 29.3  &  30.9 & 29.1  & 30.0  & 28.5  & 40.5  & 81.0\textsuperscript{$\ddag$}\\
			$\mathrm{C_{12}}$ & 15.8&   7.3  & 9.5   &  11.7 & 13.1  & 12.0  &  7.7  & 11.3  & 25.1\\
			$\mathrm{C_{33}}$ & 8.6 &   4.7  & 9.9   & 	 9.6 &  8.0  &  9.0  & 10.4  & 14.6  & 27.6\\
			$K_{I}$           &48.8 &  23.9  & 38.8  &  42.6 & 42.2  & 42.0  & 36.2  & 51.8  &105.8\\
			$K_{II}$          &17.3 &   9.4  & 19.8  &  19.2 & 16.0  & 18.0 & 20.8 & 29.1 & 55.6\textsuperscript{$\ddag$}\\
			$K_{III}$         &17.3 &   9.4  & 19.8  &  19.2 & 16.0  & 18.0  & 20.8  & 29.1  & 55.3\textsuperscript{$\ddag$}\\
			MAPE$_\textrm{LHD}$      &       & 29.589& 12.170 & 10.574 & 9.473 & 7.639 & 15.570 & 27.321 & 94.353\\
			{MAPE$^\textrm{S}_\textrm{LHD}$} & &1.678&  3.459 &  3.774 & 3.157 & 3.700 &  3.399 &  1.982 &  1.463\\
			{MAPE$^\textrm{M}_\textrm{LHD}$} &&49.210& 20.411 &	33.854 &30.520 &16.344 & 27.986 & 66.258 &140.844\\
			\hline
			$\sum$MAPE$_\textrm{7Sn}$             &       &264.400& 206.170& 200.270&251.441&148.782&271.699 &276.281 &857.523\\
			{$\sum$MAPE$^\textrm{S}_\textrm{7Sn}$}&       & 19.533&  41.948&  48.049& 92.876& 35.957& 86.296 & 12.919 & 19.157\\
			{$\sum$MAPE$^\textrm{M}_\textrm{7Sn}$}&       &413.036& 295.325& 397.200&436.740&235.242&388.655 &582.493 &1271.379\\
			\hline
		\end{tabular}
	\begin{tablenotes}[para]
		\small
		\textsuperscript{$\ddag$} {Potential does not reproduce the required isotropy of the elasticity tensor ($\mathrm{C_{11}}\neq\mathrm{C_{22}}$}~\&$/$$\wedge$~$K_{II}$$\neq$$K_{III}$). 
	\end{tablenotes}
\end{table}

\begin{table}[htbp]
	\caption{Aggregated {total} mean absolute percentage error~({MAPE}$_\textrm{A}$)\,(\%), {aggregated structural mean absolute percentage} 
		error~{MAPE}$^\textrm{S}_\textrm{A}$\,(\%), aggregated mechanical mean absolute percentage error~{MAPE}$^\textrm{M}_\textrm{A}$\,(\%), standard deviation $\sigma$.}
\label{tab:MMAPE}
\centering
\setlength{\tabcolsep}{1.0pt} 
	\begin{tabular}{c c c c c c c c c}
		\hline
		\textbf{Model}  &            \textbf{Best}     & \textbf{MACE}  & \textbf{MatterSim} & \textbf{CHGNet} & \textbf{Nequix} & \textbf{SevenNet} &\textbf{ORB} &\textbf{PET}\\
		\textbf{version}&           \textbf{Classical} &\textbf{MP-0b2} &\textbf{v1.0.0-5M} &\textbf{0.3.0} &\textbf{MP-1} &\textbf{L3I5} &\textbf{v3-C.} &\textbf{MAD}\\
		&           \textbf{Potential} &\textbf{(large)}&                   &               &              &              & \textbf{OMAT} &             \\
		\hline
		$\sum${MAPE}$_\textrm{A}$              & 153.398 & 128.163 & 132.980 & 210.580 & 117.269 & 169.717 & 171.400 & 401.684\\
		$\sigma$                                           & 34.153  &  25.543 &  23.766 &  24.084 &  17.212 &  33.412 &  34.480 & 124.583\\		
		{$\sum${MAPE}$^\textrm{S}_\textrm{A}$} &  20.415 &  28.364 &  26.706 &  56.316 &  31.806 &  25.687 &  11.024 &  20.911\\
		$\sigma$                                           &   2.294 &  5.782  &   6.569 &  12.426 &   4.709 &   5.072 &   1.655 &   3.463\\
		{$\sum${MAPE}$^\textrm{M}_\textrm{A}$} & 269.989 & 203.545 & 278.844 & 358.471 & 204.951 & 275.318 & 381.365 & 608.667\\
		$\sigma$                                           &  49.080 &  33.975 &  45.368 &  43.640 &  26.407 &  46.471 &  70.089 & 182.754\\
		\hline
	\end{tabular}
\begin{tablenotes}[para]
	\small
\end{tablenotes}
\end{table}

Now, let's consider the cost of using classical potentials and uMLIPs in molecular calculations, see Table~\ref{tab:Mtime}.
A comparison of the intrinsic computational complexity of the potentials themselves on a single CPU core, without hardware interference, seems to be the most reliable because it measures pure sequential complexity.
First, let's consider a structure consisting of a single type of atom, namely Si. The same is true for Ge and Sn. We can see that the computational cost for uMLIPs here is 10$^3$?10$^5$ higher than for classical potentials. The lowest overhead is for \textbf{MatterSim} (990), and the highest is for the large model of \textbf{Mace} (65,000). A 6.5-fold difference in overhead is seen between the large and small \textbf{MACE} models. For the two-element structure MoS$_2$, the overhead approximately doubles, and \textbf{REBO} is 13 times slower than the classical Stillinger?Weber (\textbf{SW}) potential. 
A difference in the same order of magnitude, ranging from 10$^4$ to 10$^5$, was observed between the classical \textbf{EAM} potential and the \textbf{MACE-MP-0} uMLIP for cubic tungsten on a single CPU \cite{Shuang2025}. For 3D MoS$_2$, \textbf{MACE-MP-0} and \textbf{CHGNet} uMLIPs were found to be more than 10$^4$ times slower than classical potentials when running on a 32-core processor \cite{Bilgili2026}. 
It is worth noting that the situation is similar for GPUs, and on a single NVIDIA A100 GPU, the classic \textbf{EAM} is approximately 23,000 times faster (more than 4 orders of magnitude) than the standard \textbf{MACE} in LAMMPS \cite{Phuthi2026}.
We should also keep in mind the memory requirements of uMLIPs. The author of \textbf{CHGNet} uMLIP \cite{deng2023chgnet} reported that memory bottlenecks occur when the number of atoms ranges from 6,250 to 14,700, depending on the hardware. Furthermore, based on his calculations, a single NVIDIA H100 GPU is estimated to be equivalent to approximately 500 CPU cores of an AMD EPYC 9554 processor for MD computations.
For \textbf{GRACE-2L-L}, uMLIP on an NVIDIA A100-80GB GPU could only accommodate 20,000 atoms. Optimized GRACE models in LAMMPS run 2?5 times faster than those run in the ASE environment \cite{Lysogorskiy2026}. 

\begin{table}[htbp]
\caption{Relative performance measured as normalized seconds/timestep in molecular dynamics (MD) simulation.}
\label{tab:Mtime}
\centering
\setlength{\tabcolsep}{1.0pt} 
	\begin{tabular}{c c c c c c c c c c c c}
		\hline \multicolumn{12}{c}{MoS$_2$} \\ 
		\hline 
		\textbf{Model} &\textbf{SW}   & \textbf{REBO} & \textbf{MACE} & \textbf{MACE} & \textbf{MACE} & \textbf{MatterSim} & \textbf{CHGNet} & \textbf{Nequix} & \textbf{SevenNet} &\textbf{ORB} &\textbf{PET}\\
		\textbf{version}&\textbf{2016}&               &\textbf{MP-0b2} &\textbf{MPA-0} &\textbf{MP-0b2} &\textbf{v1.0.0-5M} &\textbf{0.3.0} &\textbf{MP-1} &\textbf{L3I5} &\textbf{v3-C.} &\textbf{MAD}\\
		&             &               &\textbf{(small)}&\textbf{(medium)} &\textbf{(large)}&                   &               &              &              & \textbf{OMAT} &             \\
		\hdashline  
		s/timestep & 1 & 13 & 19526 & 52596 & 120777 & 2495 & 4924 & 3090 & 8838 & 29917 & 5603\\
		
		\hline \multicolumn{12}{c}{Si} \\ 
		\hline
		\textbf{Model} &\textbf{SW}   & \textbf{MEAM} & \textbf{MACE} & \textbf{MACE} & \textbf{MACE} & \textbf{MatterSim} & \textbf{CHGNet} & \textbf{Nequix} & \textbf{SevenNet} &\textbf{ORB} &\textbf{PET}\\
		\textbf{version}&\textbf{1985}& \textbf{2011} &\textbf{MP-0b2} &\textbf{MPA-0} &\textbf{MP-0b2} &\textbf{v1.0.0-5M} &\textbf{0.3.0} &\textbf{MP-1} &\textbf{L3I5} &\textbf{v3-C.} &\textbf{MAD}\\
		&             &               &\textbf{(small)}&\textbf{(medium)} &\textbf{(large)}&              &               &              &              & \textbf{OMAT} &             \\
		\hdashline
		s/timestep & 1 & 2 & 10000 & 28000 & 65000& 990 & 1463 & 1578 & 3367 & 11447 & 2562 \\
		\hline
	\end{tabular}
\end{table}

\clearpage
\section{Conclusions}
\label{sec:Conc}

This article compares the ability of various universal machine learning interatomic potentials (uMLIPs) and the best classical potentials to reproduce the structural and mechanical properties of four two-dimensional materials (MoS$_2$, Si, Ge, and Sn) across their multiple phases. Predictions from DFT and molecular calculations were compared for the interatomic potentials listed in Section~\ref{sec:Ip}.
The mean absolute percentage error (MAPE) metric was used here. A comparison of the computational costs of the potentials under study was also conducted.

{\begin{itemize}
		\item Of the uMLIPs, only two, \textbf{MACE} and \textbf{MatterSim}, never returned non-physical results for all 19 structures; 
		
		\item The \textbf{Nequix} uMLIP had a lower aggregated total mean absolute percentage error, but it could return non-physical results;
		
		\item Most uMLIPs underperform compared to classical potentials and often struggle to satisfy Neumann?s principle;
		
		\item The uMLIPs analyzed are three to five orders of magnitude more computationally expensive than classical potentials, as shown in Table~\ref{tab:Mtime};
		
		\item The very high memory requirements of uMLIPs severely limit their use in large-scale molecular dynamics (MD) simulations;
		
		\item When it comes to performance, accuracy, and cost, classical potentials are still the best option for molecular calculations. 
\end{itemize} }

The universal machine learning interatomic potentials (uMLIPs) have been significantly improved in terms of accuracy and performance (MAPE) compared to the previously developed specialized machine learning interatomic potentials (sMLIPs) \cite{Mazdziarz2021, Mazdziarz2023, Mazdziarz2023b, Mazdziarz2025}. The uMLIPs, like the earlier sMLIPs, still often yield non-physical predictions that manifest as violations of the principle of symmetry (Neumann's principle). However, they do not demonstrate any particular superiority compared to the best classical models in the cases analyzed here, and their computational cost and memory requirements severely limit their use.

\clearpage
\newpage
\section*{Supporting Information}
\label{sec:SuppI}

\subsection{Interatomic Potentials}
\label{sec:SIIP}

The parametrizations of the classical potentials enumerated hereafter were obtained from the~\href{https://www.ctcms.nist.gov/potentials/}{NIST} Interatomic Potentials Repository \cite{NIST} and/or from~LAMMPS code sources and/or through the assistance of the authors of the aforementioned publications. Access to all uMLIPs was made possible by the uMLIP-Interactive tool \cite{LEBEDA20266766}. In LAMMPS on a CPU, the MACE uMLIP uses the interatomic potential style (pair\_style mace) to run \href{https://mace-docs.readthedocs.io/en/latest/index.html}{MACE} machine learning models. 


\begin{enumerate}
	\item \textbf{MACE-MP-0b2 (large)} \cite{Batatia2022mace}: {{MACE} is Higher Order Equivariant Message Passing Neural Networks for Fast and Accurate Force Fields}, using MPTrj Training Set, large variant
	\begin{itemize}
		\item \textbf{MACE-MP-0b2 (small)} \cite{Batatia2022mace}: {{MACE}}, using MPTrj Training Set, small variant
		\item \textbf{MACE-MPA-0 (medium)} \cite{Batatia2022mace}:  {{MACE}}, using MPTrj + sAlex (Subsampled version of the Alexandria dataset) Training Set, medium variant
		\item \textbf{MACE-MATPES-r2SCAN-0 (medium)} \cite{Batatia2022mace}: {{MACE}}, using MATPES-r2SCAN Training Set, medium variant
	\end{itemize}
	\item \textbf{MatterSim-v1.0.0-5M} \cite{yang2024mattersim}: {MatterSim} is models based on the M3GNet architecture, using MatterSim Training Set, large variant
	\item \textbf{CHGNet-0.3.0} \cite{deng2023chgnet}: {CHGNet} is pretrained universal neural network potential for charge-informed atomistic modeling, using MPTrj Training Set, default variant
	\item \textbf{Nequix-MP-1} \cite{koker2025trainingfoundationmodelmaterials}: {Nequix} is a compact E(3)-equivariant potential, using MPTrj Training Set, default variant
	\item \textbf{SevenNet-L3I5} \cite{Park2024}: {SevenNet} is a graph neural network (GNN)-based interatomic potential based on {Nequix}, using MPTrj Training Set, improved variant  
	\begin{itemize}
		\item \textbf{SevenNet-0} \cite{Park2024}: {SevenNet}, using MPTrj Training Set, default variant 
	\end{itemize}
	\item \textbf{ORB-v3 Conservative OMAT} \cite{rhodes2025orbv3atomisticsimulationscale}: {ORB} is a Graph Network Simulator (GNS) augmented with	a smoothed graph attention mechanism, using MPtrj+Alex+OMat24 Training Set \cite{Jain2013,Cavignac_2026,barroso_omat24}, improved variant
	\item \textbf{PET-MAD} \cite{PETMAD2025}: {PET-MAD} is model based on the Point Edge Transformer (PET) architecture, using OMat24 Training Set, default variant 
	
	\noindent\dotfill
	\item \textbf{SW2016-MoS$_2$} \cite{Kandemir_2016}: {SW} is the Stillinger-Weber potential fitted to lattice parameters, distance
	between two chalcogen atoms and elastic constants for SL 1H-MoS$_2$ obtained from DFT calculations
	\item \textbf{REBO-MoS$_2$} \cite{Liang2009,Liang2012}: {REBO} is the reactive many-body potential fitted to structure and energetics of Mo molecules, three-dimensional Mo crystals, two-dimensional Mo structures, small S molecules, and binary Mo-S crystal structures 
	\item \textbf{MEAM2011-Si} \cite{Du2011}: {MEAM} is spline-based modified embedded-atom method potential for Si fitted for silicon interstitials
	\item \textbf{SW1985-Si} \cite{Stillinger1985}: {SW} is the Stillinger--Weber potential fitted to solid and liquid forms of Si
	\item \textbf{SW1986-Ge} \cite{Ding1986}: {SW} is the Stillinger-Weber potential fitted to cohesive energy and elastic constants of {3D} germanium
	\item \textbf{MEAM2018b-Sn} \cite{Ko2018}: {MEAM} is the modified embedded atom method potential whose parameters were optimized based on the force-matching method using the density functional theory (DFT) database of energies and forces of atomic configurations of tin under different conditions 
\end{enumerate}

\newpage

\subsection{Results}\label{SIR}

\begin{table}[!h]
	\caption{Structural and mechanical properties of SL 1H-MoS$_2$ from molecular calculations: lattice parameters a, b\,(\AA), \comm{and $\gamma$\,($^\circ$)} average cohesive energy $E_{c}$\,(eV/atom), average bond length $d$\,(\AA), average height $h$\,(\AA), 2D elastic constants $\mathrm{C_{ij}}$\,(N/m), 2D Kelvin moduli $K_i$\,(N/m), {total} mean absolute percentage error~(MAPE)\,(\%), {structural mean absolute percentage error~(MAPE$^\textrm{S}$)\,(\%), mechanical mean absolute percentage error~(MAPE$^\textrm{M}$)\,(\%)}.\comm{, relative performance measured as normalized timesteps/second in molecular dynamics (MD) simulation.}}
	\label{tab:MoS2_H}
	\centering
	\setlength{\tabcolsep}{1.0pt} 
	\tiny
		\begin{tabular}{c c c c c c c c c c c c c c}
			\hline
			\textbf{Model} &\textbf{DFT}& \textbf{REBO} & \textbf{MACE} & \textbf{MACE} & \textbf{MACE} & \textbf{MACE} & \textbf{MatterSim} & \textbf{CHGNet} & \textbf{Nequix} & \textbf{SevenNet} & \textbf{SevenNet} &\textbf{ORB} &\textbf{PET}\\
			\textbf{version}& &  &\textbf{MP-0b2} &\textbf{MPA-0} &\textbf{MP-0b2} &\textbf{MATPES} &\textbf{v1.0.0-5M} &\textbf{0.3.0} &\textbf{MP-1} &\textbf{L3I5} &\textbf{0} &\textbf{v3-C.} &\textbf{MAD}\\
			                & &  &\textbf{(small)}&\textbf{(medium)} &\textbf{(large)} &\textbf{r2SCAN-0}&              &               &              &              &           & \textbf{OMAT}\\
			\hline      a   & 3.165 & 3.168 & 3.194 & 3.215 & 3.194 & 3.197 & 3.190 & 3.193 & 3.187 & 3.190 & 3.192 & 3.182 & 3.167\\
			            b   & 3.165 & 3.168 & 3.194 & 3.215 & 3.194 & 3.197 & 3.190 & 3.193 & 3.187 & 3.190 & 3.192 & 3.182 & 3.167\\
			$-E_{c}$        & 5.641 & 7.162	& 5.103 & 5.099 & 5.104 & 4.754 & 5.004 & 5.281 & 5.010 & 5.166 & 5.138 & 3.994 & 6.782\\
			$d$             & 2.403 & 2.445	& 2.419	& 2.424 & 2.433 & 2.412 & 2.419 & 2.419 & 2.419 & 2.419 & 2.419 & 2.412 & 2.396\\
			$h$             & 3.120 & 3.242 & 3.125 & 3.114 & 3.128 & 3.094 & 3.132 & 3.125 & 3.133 & 3.133 & 3.130 & 3.123 & 3.093\\
			$\mathrm{C_{11}}$ & 126.5 & 154.4 & 151.2 & 179.7 & 151.7 & 167.8 & 164.6 & 140.4 & 128.9 & 140.7 & 137.7 & 144.5\textsuperscript{$\ddag$} & 175.6\textsuperscript{$\ddag$}\\
			$\mathrm{C_{22}}$ & 126.5 & 154.4 & 151.2 & 179.7 & 151.7 & 167.8 & 164.6 & 140.4 & 128.9 & 140.7 & 137.7 & 144.6\textsuperscript{$\ddag$} & 176.1\textsuperscript{$\ddag$}\\
			$\mathrm{C_{12}}$ & 28.5  & 45.8  & 44.8  & 56.5  & 53.9  &  48.9 &  42.5 &  81.5 &  61.2 &  63.4 &  67.6 &  51.9 &  76.3\\
			$\mathrm{C_{33}}$ & 49.0  & 54.3  &  53.2 &	61.6  & 48.9  &  59.4 &  61.0 &  29.6 &  33.9 &  38.6 &  35.1 &  46.3 &  50.2\\
			$K_{I}$           & 155.0 & 200.2 & 196.0 & 236.3 & 205.6 & 216.7 & 207.1 & 221.9 & 190.1 & 204.2 & 205.2 & 196.4 & 252.2\\
			$K_{II}$          & 98.0  & 108.6 & 106.4 & 123.2 &  97.9 & 118.9 & 122.0 &  59.0 &  67.8 &  77.3 &  70.1 &  92.7\textsuperscript{$\ddag$} &  99.5\textsuperscript{$\ddag$}\\
			$K_{III}$         & 98.0  & 108.6 & 106.4 & 123.2 &  97.9 & 118.9 & 122.0 &  59.1 &  67.8 &  77.3 &  70.1 &  92.5\textsuperscript{$\ddag$} & 100.4\textsuperscript{$\ddag$}\\
			{MAPE$_\textrm{H}$}            & & 16.602 & 13.384 & 27.149 & 14.560 & 21.628 & 19.203 & 31.596 & 20.611 & 20.941 & 23.670 & 15.381 & 28.037\\
			{MAPE$^\textrm{S}_\textrm{H}$} & & 1.949  & 0.878  &  1.397 &  1.102 &  1.079 &  0.863 &  0.865 &  0.819 &  0.867 &  0.896 &  0.499 &  0.420\\
			{MAPE$^\textrm{M}_\textrm{H}$} & & 44.401 & 26.896 & 47.404 & 28.561 & 43.833 & 36.954 & 51.857 & 38.993 & 37.397 & 41.861 & 44.788 & 57.064\\
			\hline
		\end{tabular}
	{\begin{tablenotes}[para,flushleft]
			\raggedright
			\small
			\textsuperscript{$\ddag$} {Potential does not reproduce the required isotropy of the elasticity tensor ($\mathrm{C_{11}}\neq\mathrm{C_{22}}$}~\&$/$$\wedge$~$K_{II}$$\neq$$K_{III}$). 
	\end{tablenotes} }
\end{table}

\begin{table}[!h]
	\caption{Structural and mechanical properties of SL 1T-MoS$_2$ from molecular calculations: lattice parameters a, b\,(\AA), \comm{and $\gamma$\,($^\circ$)} average cohesive energy $E_{c}$\,(eV/atom), average bond length $d$\,(\AA), average height $h$\,(\AA), 2D elastic constants $\mathrm{C_{ij}}$\,(N/m), 2D Kelvin moduli $K_i$\,(N/m), {total} mean absolute percentage error~(MAPE)\,(\%), {structural mean absolute percentage error~(MAPE$^\textrm{S}$)\,(\%), mechanical mean absolute percentage error~(MAPE$^\textrm{M}$)\,(\%)}.\comm{, relative performance measured as normalized timesteps/second in molecular dynamics (MD) simulation.}}
	\label{tab:MoS2_T}
	\centering
	\setlength{\tabcolsep}{1.0pt} 
	\tiny
		\begin{tabular}{c c c c c c c c c c c c c c}
			\hline
			\textbf{Model} &\textbf{DFT}& \textbf{REBO} & \textbf{MACE} & \textbf{MACE} & \textbf{MACE} & \textbf{MACE} & \textbf{MatterSim} & \textbf{CHGNet} & \textbf{Nequix} & \textbf{SevenNet} & \textbf{SevenNet} &\textbf{ORB} &\textbf{PET}\\
			\textbf{version}& &  &\textbf{MP-0b2} &\textbf{MPA-0} &\textbf{MP-0b2} &\textbf{MATPES} &\textbf{v1.0.0-5M} &\textbf{0.3.0} &\textbf{MP-1} &\textbf{L3I5} &\textbf{0} &\textbf{v3-C.} &\textbf{MAD}\\
			& &  &\textbf{(small)}&\textbf{(medium)} &\textbf{(large)} &\textbf{r2SCAN-0}&              &               &              &              &           & \textbf{OMAT}\\
			\hline      a   & 3.194 & 3.194 & 3.197 & 3.208 & 3.211 & 3.187 & 3.188 & 3.187 & 3.193 & 3.197 & 3.190 & 3.190 & 3.100\\
			            b   & 3.194 & 3.194 & 3.197 & 3.208 & 3.211 & 3.187 & 3.188 & 3.187 & 3.193 & 3.197 & 3.190 & 3.190 & 3.100\\
			$-E_{c}$        & 5.521 & 7.047 & 4.819 & 4.822 & 4.851 & 4.463 & 4.728 & 5.082 & 4.754 & 4.904 & 5.078 & 3.710 & 6.512\\
			$d$             & 2.422 & 2.445 & 2.433 & 2.438 & 2.433 & 2.419 & 2.433 & 2.440 & 2.433 & 2.440 & 2.419 & 2.426 & 2.405\\
			$h$             & 3.142 & 3.211 & 3.163 & 3.171 & 3.172 & 3.137 & 3.190 & 3.193 & 3.175& 3.182 & 3.144 & 3.162 & 3.208\\
			$\mathrm{C_{11}}$ & 84.1 & 118.2 & 94.0 & 133.2 & 97.0  & 139.4 & 130.8 & 130.7 & 93.8 & 121.8 & 129.0 & 120.9 & 159.2\textsuperscript{$\ddag$}\\
			$\mathrm{C_{22}}$ & 84.1 & 118.2 & 94.0 & 133.2 & 97.0  & 139.4 & 130.8 & 130.7 & 93.8 & 121.8 & 129.0 & 120.9 & 161.0\textsuperscript{$\ddag$}\\
			$\mathrm{C_{12}}$ &  5.0 &  32.4 & 40.9 &  55.4 & 33.3  &  27.9 &  44.3 &  57.3 & 35.9 &  48.6 &  56.5 &  42.3 &  35.3\\
			$\mathrm{C_{33}}$ & 39.6 &  42.9 & 26.6 &  38.9 & 31.9  &  55.7 &  43.3 &  36.7 & 29.0 &  36.6 &  36.2 &  39.2 &  62.7\\
			$K_{I}$           & 89.1 & 150.6 & 134.9& 188.6 &130.4  & 167.4 & 175.2 & 187.9 &129.7 & 170.4 & 185.5 & 163.3 & 195.4\\
			$K_{II}$          & 79.2 &  85.8 &  53.2&  77.8 & 63.7  & 111.5 &  86.5 &  73.4 & 57.9\textsuperscript{$\ddag$} &  73.2 &  72.5 &  78.6\textsuperscript{$\ddag$} &  124.8\textsuperscript{$\ddag$}\\
			$K_{III}$         & 79.2 &  85.8 &  53.2&  77.8 & 63.7  & 111.5 &  86.5 &  73.4 & 58.0\textsuperscript{$\ddag$} &  73.2 &  72.5 &  78.3\textsuperscript{$\ddag$} &  125.3\textsuperscript{$\ddag$}\\
			{MAPE$_\textrm{H}$}            & &62.843 & 75.427&104.754& 59.706& 68.367& 86.545&108.297& 65.172& 90.792&106.636& 79.538& 92.205\\
			{MAPE$^\textrm{S}_\textrm{H}$} & & 0.786 &  0.323&  0.616&  0.606&  0.186&  0.586&  0.699&  0.395&  0.534&  0.125&  0.264&  2.180\\
			{MAPE$^\textrm{M}_\textrm{H}$} & &114.575&122.518&166.318& 98.353&116.823&140.297&168.063&107.975&144.306&165.914&143.773&150.675\\
			\hline
		\end{tabular}
	{\begin{tablenotes}[para,flushleft]
			\raggedright
			\small
			\textsuperscript{$\ddag$} {Potential does not reproduce the required isotropy of the elasticity tensor ($\mathrm{C_{11}}\neq\mathrm{C_{22}}$}~\&$/$$\wedge$~$K_{II}$$\neq$$K_{III}$). 
	\end{tablenotes} }
\end{table}

\begin{table}[!h]
	\caption{Structural and mechanical properties of SL 1T'-MoS$_2$ from molecular calculations: lattice parameters a, b\,(\AA), \comm{and $\gamma$\,($^\circ$)} average cohesive energy $E_{c}$\,(eV/atom), average bond length $d$\,(\AA), average height $h$\,(\AA), 2D elastic constants $\mathrm{C_{ij}}$\,(N/m), 2D Kelvin moduli $K_i$\,(N/m), {total} mean absolute percentage error~(MAPE)\,(\%), {structural mean absolute percentage error~(MAPE$^\textrm{S}$)\,(\%), mechanical mean absolute percentage error~(MAPE$^\textrm{M}$)\,(\%)}.\comm{, relative performance measured as normalized timesteps/second in molecular dynamics (MD) simulation.}}
	\label{tab:MoS2_T'}
	\centering
	\setlength{\tabcolsep}{0.50pt} 
	\tiny
		\begin{tabular}{c c c c c c c c c c c c c c}
			\hline
			\textbf{Model} &\textbf{DFT}& \textbf{REBO} & \textbf{MACE} & \textbf{MACE} & \textbf{MACE} & \textbf{MACE} & \textbf{MatterSim} & \textbf{CHGNet} & \textbf{Nequix} & \textbf{SevenNet} & \textbf{SevenNet} &\textbf{ORB} &\textbf{PET}\\
			\textbf{version}& &  &\textbf{MP-0b2} &\textbf{MPA-0} &\textbf{MP-0b2} &\textbf{MATPES} &\textbf{v1.0.0-5M} &\textbf{0.3.0} &\textbf{MP-1} &\textbf{L3I5} &\textbf{0} &\textbf{v3-C.} &\textbf{MAD}\\
			& &  &\textbf{(small)}&\textbf{(medium)} &\textbf{(large)} &\textbf{r2SCAN-0}&              &               &              &              &           & \textbf{OMAT}\\
			\hline      a   & 5.751 & 5.563 & 5.817 & 5.854 & 5.939 & 5.762 & 5.870 & 5.763 & 5.918 & 5.686	& 5.525 & 5.755 & 5.799\\
			            b   & 3.177 & 3.245 & 3.179 & 3.194 & 3.222 & 3.145 & 3.149 & 3.215 & 3.178 & 3.212 & 3.189 & 3.172 & 3.143\\
			$-E_{c}$        & 5.562 & 6.934 & 4.887	& 4.947 & 4.887 & 4.557 & 4.779 & 5.119 & 4.807 & 4.977 & 5.078 & 3.798 & 6.585\\
			$d$             & 2.415 & 2.468 & 2.440 & 2.427 & 2.456 & 2.430 & 2.435 & 2.429 & 2.448 & 2.459 & 2.419 & 2.438 & 2.423\\
			$h$             & 3.364 & 3.781 & 3.406 & 3.407 & 3.330 & 3.382 & 3.378 & 3.436 & 3.269 & 3.436 & 3.145 & 3.457 & 3.375\\
			$\mathrm{C_{11}}$ & 68.1 & 56.8 & 117.6 &  141.0& 105.0 & 154.5 & 130.2 & 125.6 & 117.9 & 134.2	& 128.4 & 147.2 & 153.5\\
			$\mathrm{C_{22}}$ & 78.9 & 113.0& 90.7  &  132.1&  92.0 & 153.8 & 122.2 & 126.2 & 105.3 & 118.1	& 129.1 & 121.1 & 170.0\\
			$\mathrm{C_{12}}$ & 18.2 & 23.1 & 46.2  &   66.3&  49.6 &  45.7 &  45.8 &  55.6 &  45.1 &  55.8	&  56.9 &  53.5 &  52.5\\
			$\mathrm{C_{33}}$ & 43.2 & 70.5 & 29.4  &   49.2&  32.8 &  60.2 &  45.5 &  35.8 &  33.3 &  42.6 &  36.6 &  56.7 &  32.7\\
			$K_{I}$           & 90.9 &121.3 & 148.3 &  202.7& 147.7 & 199.8 & 171.8 & 181.6 & 156.3 & 181.3	& 185.6 & 186.0 & 213.6\\
			$K_{II}$          & 56.1 & 48.5 &  60.0 &   70.4&  49.3 & 108.5 &  80.5 &  70.3 &  66.9 &  71.0 &  71.9 &  82.2 & 109.8\\
			$K_{III}$         & 86.4 &141.0 &  58.7	&   98.4&  65.5 & 120.3 &  91.1 &  71.5 &  66.6 &  85.1	&  73.1 & 113.4 &  65.3\\
			MAPE$_\textrm{T'}$              & & 25.395 & 32.600 & 52.512 & 32.169 & 57.055 & 38.280 & 43.470 & 34.342 & 41.573 & 45.615& 51.069 & 60.837\\
			{MAPE$^\textrm{S}_\textrm{T'}$} & & 5.000  & 0.872  &  1.027 &  1.840 &  0.587 &  1.040 &  1.033 &  1.783 &  1.557 &  2.740&  0.984 &  0.646\\
			{MAPE$^\textrm{M}_\textrm{T'}$} & & 54.073 & 57.547 & 86.527 & 56.414 & 98.825 & 67.439 & 70.639 &	60.785 & 69.451 & 73.558& 99.883 & 104.756\\
			\hline
			$\sum$MAPE$_\textrm{3MoS$_2$}$             & & 104.840 & 121.411 & 184.415 & 106.435 & 147.050 & 144.024 & 183.360 & 120.122 & 153.305 & 175.921 & 145.986 & 181.081\\
			{$\sum$MAPE$^\textrm{S}_\textrm{3MoS$_2$}$}& &   7.736 &   2.074 &   3.040 &   3.549 &   1.852 &   2.490 &   2.598 &   2.996 &   2.959 &   3.760 &   1.746 &   3.246\\
			{$\sum$MAPE$^\textrm{M}_\textrm{3MoS$_2$}$}& & 213.100 & 206.960 & 300.248 & 183.328 & 259.481 & 244.662 & 290.530 & 207.725 & 251.154 & 281.332 & 288.422 & 312.532\\
			
			\hline
		\end{tabular}
	{\begin{tablenotes}[para,flushleft]
			\raggedright
			\small
	\end{tablenotes} }
\end{table}
\begin{table}[!h]
	\caption{Structural and mechanical properties of flat (F) silicene from molecular calculations: lattice parameters a, b\,(\AA), \comm{and $\gamma$\,($^\circ$)} average cohesive energy $E_{c}$\,(eV/atom), average bond length $d$\,(\AA), average height $h$\,(\AA), 2D elastic constants $\mathrm{C_{ij}}$\,(N/m), 2D Kelvin moduli $K_i$\,(N/m), {total} mean absolute percentage error~(MAPE)\,(\%), {structural mean absolute percentage error~(MAPE$^\textrm{S}$)\,(\%), mechanical mean absolute percentage error~(MAPE$^\textrm{M}$)\,(\%)}. \comm{, relative performance measured as normalized timesteps/second in molecular dynamics (MD) simulation.}}
	\label{tab:Si_F}
	\centering
	\setlength{\tabcolsep}{1.0pt} 
	\scriptsize
		\begin{tabular}{c c c c c c c c c c}
			\hline
			\textbf{Model} &\textbf{DFT}& \textbf{MEAM}  & \textbf{MACE} & \textbf{MatterSim} & \textbf{CHGNet} & \textbf{Nequix} & \textbf{SevenNet} &\textbf{ORB} &\textbf{PET}\\
			\textbf{version}&           & \textbf{2011}  &\textbf{MP-0b2} &\textbf{v1.0.0-5M} &\textbf{0.3.0} &\textbf{MP-1} &\textbf{L3I5} &\textbf{v3-C.} &\textbf{MAD}\\
							&           &                &\textbf{(large)}&                   &               &              &              & \textbf{OMAT} &             \\
			\hline      a   & 3.855 & 3.960 & 3.881 & 3.934 & 3.969 & 3.904 & 3.916 & 3.914 & 3.828\\
			            b   & 3.855 & 3.960 & 3.881 & 3.934 & 3.969 & 3.904 & 3.916 & 3.914 & 3.828\\
			$-E_{c}$        & 4.562 & 3.793 & 3.871 & 3.289 & 3.505 & 3.651 & 3.795 & 3.027 & 4.207\\
			$d$             & 2.225 & 2.288 & 2.244	& 2.272 & 2.293 & 2.258 & 2.265 & 2.275 & 2.209\\
			$h$             & 0.0 & 0.0 & 0.0 & 0.0 & 0.0 & 0.0 & 0.0 & 0.0 & 0.0\\
			$\mathrm{C_{11}}$ & 84.8 & 84.0 & 74.0 & 71.7 & 119.6\textsuperscript{$\ddag$} & 104.4 & 108.0 & 103.7 & 124.2\textsuperscript{$\ddag$}\\
			$\mathrm{C_{22}}$ & 84.8 & 84.0 & 74.0 & 71.7 & 119.2\textsuperscript{$\ddag$} & 104.4 & 108.0 & 103.7 & 122.7\textsuperscript{$\ddag$}\\
			$\mathrm{C_{12}}$ & 40.6 & 40.8 & 29.4 & 22.6 &  57.2 & 30.4 & 28.1 & 28.5 & 46.8\\
			$\mathrm{C_{33}}$ & 22.1 & 21.6 & 22.3 & 24.5 &  31.6 & 37.0 & 40.0 & 37.5 & 38.7\\
			$K_{I}$           & 125.4& 124.8& 103.5& 94.4 & 176.6 &134.8 &136.1 &132.2 &170.3\\
			$K_{II}$          & 44.3 & 43.3 &  44.6& 49.1 &  62.2\textsuperscript{$\ddag$} & 74.0\textsuperscript{$\ddag$} & 80.0 & 75.2\textsuperscript{$\ddag$} & 76.6\textsuperscript{$\ddag$}\\
			$K_{III}$         & 44.3 & 43.3 &  44.6& 49.1 &  63.2\textsuperscript{$\ddag$} & 74.1\textsuperscript{$\ddag$} & 80.0 & 75.0\textsuperscript{$\ddag$} & 77.5\textsuperscript{$\ddag$}\\
			MAPE$_\textrm{F}$              & & 3.147 & 8.187 & 15.151 & 29.255 & 27.702 & 32.554 & 29.803 & 34.145\\
			{MAPE$^\textrm{S}_\textrm{F}$} & & 2.759 & 0.733 &  2.076 &  2.982 &  1.344 &  1.657 &  1.782 &  0.711\\
			{MAPE$^\textrm{M}_\textrm{F}$} & &15.932 &22.339 & 40.973 & 56.486 & 52.563 & 56.739 & 65.542 & 52.521\\
			\hline
		\end{tabular}
	\begin{tablenotes}[para,flushleft]
		\raggedright
		\small
		\textsuperscript{$\ddag$} {Potential does not reproduce the required isotropy of the elasticity tensor ($\mathrm{C_{11}}\neq\mathrm{C_{22}}$}~\&$/$$\wedge$~$K_{II}$$\neq$$K_{III}$). 
	\end{tablenotes}
\end{table}

\begin{table}[!h]
	\caption{Structural and mechanical properties of low-buckled (LB) silicene from molecular calculations: lattice parameters a, b\,(\AA), \comm{and $\gamma$\,($^\circ$)} average cohesive energy $E_{c}$\,(eV/atom), average bond length $d$\,(\AA), average height $h$\,(\AA), 2D elastic constants $\mathrm{C_{ij}}$\,(N/m), 2D Kelvin moduli $K_i$\,(N/m), {total} mean absolute percentage error~(MAPE)\,(\%), {structural mean absolute percentage error~(MAPE$^\textrm{S}$)\,(\%), mechanical mean absolute percentage error~(MAPE$^\textrm{M}$)\,(\%)}. \comm{, relative performance measured as normalized timesteps/second in molecular dynamics (MD) simulation.}}
	\label{tab:Si_LB}
	\centering
	\setlength{\tabcolsep}{1.0pt} 
	\scriptsize
		\begin{tabular}{c c c c c c c c c c}
			\hline
			\textbf{Model} &\textbf{DFT}& \textbf{MEAM}  & \textbf{MACE} & \textbf{MatterSim} & \textbf{CHGNet} & \textbf{Nequix} & \textbf{SevenNet} &\textbf{ORB} &\textbf{PET}\\
			\textbf{version}&           & \textbf{2011}  &\textbf{MP-0b2} &\textbf{v1.0.0-5M} &\textbf{0.3.0} &\textbf{MP-1} &\textbf{L3I5} &\textbf{v3-C.} &\textbf{MAD}\\
							&           &                &\textbf{(large)}&                   &               &              &              & \textbf{OMAT} &             \\
			\hline      a   & 3.828 & 3.837 & 3.720 & 3.841 & 3.800 & 3.904 & 3.907 & 3.854 & 3.830\\
			            b   & 3.828 & 3.837 & 3.720 & 3.841 & 3.800 & 3.904 & 3.907 & 3.854 & 3.830\\
			$-E_{c}$        & 4.577 & 3.851 & 4.004 & 3.327 & 3.607 & 3.651 & 3.795 & 3.050 & 4.207\\
			$d$             & 2.249 & 2.297 & 2.314 & 2.279 & 2.391 & 2.251 & 2.265 & 2.275 & 2.223\\
			$h$             & 0.421 & 0.608 & 0.854 & 0.535 & 0.956 & 0.000\textsuperscript{$\star$} & 0.191 & 0.509 & 0.000\textsuperscript{$\star$}\\
			$\mathrm{C_{11}}$ & 69.2 & 47.1 & 72.9 & 75.9 & 114.0\textsuperscript{$\ddag$} & 104.5 & 108.4 & 89.4 & 122.9\textsuperscript{$\ddag$}\\
			$\mathrm{C_{22}}$ & 69.2 & 47.1 & 72.9 & 75.9 & 111.8\textsuperscript{$\ddag$} & 104.5 & 108.4 & 89.4 & 118.7\textsuperscript{$\ddag$}\\
			$\mathrm{C_{12}}$ & 22.1 & 10.6 & 19.0 & 21.7 & -43.6 &  30.4 &  28.2 & 19.7 & 46.6\\
			$\mathrm{C_{33}}$ & 23.6 & 18.3 & 27.0 & 27.1 &  76.6 &  37.0 &  40.1 & 34.8 & 38.4\\
			$K_{I}$           & 91.3 & 57.7 & 91.9 & 97.6 & 156.4 & 134.8 & 136.6 &109.1 &167.4\\
			$K_{II}$          & 47.1 & 36.5 & 53.9 & 54.2 &  69.3\textsuperscript{$\ddag$} &  74.1 &  80.1 & 69.8 & 74.2\textsuperscript{$\ddag$}\\
			$K_{III}$         & 47.1 & 36.5 & 53.9 & 54.2 & 153.2\textsuperscript{$\ddag$} &  74.1 &  80.1 & 69.6 & 76.7\textsuperscript{$\ddag$}\\
			MAPE$_\textrm{LB}$              & & 23.569 & 16.012 & 10.807 & 95.684 & 40.257 & 39.771 & 24.081 & 53.033\\
			{MAPE$^\textrm{S}_\textrm{LB}$} & & 11.741 & 27.792 &  7.277 & 33.711 &	26.018 & 14.870 &  5.859 & 25.317\\
			{MAPE$^\textrm{M}_\textrm{LB}$} & & 41.384 & 19.513 & 33.051 &142.571 &	62.552 & 65.041 & 58.216 & 72.958\\
			\hline
		\end{tabular}
	\begin{tablenotes}[para,flushleft]
		\raggedright
		\small
		\textsuperscript{$\star$} Potential reproduces a flat structure, not a low-buckled one, \\
		\textsuperscript{$\ddag$} {Potential does not reproduce the required isotropy of the elasticity tensor ($\mathrm{C_{11}}\neq\mathrm{C_{22}}$}~\&$/$$\wedge$~$K_{II}$$\neq$$K_{III}$). 
	\end{tablenotes}
\end{table}

\begin{table}[!h]
	\caption{Structural and mechanical properties of trigonal dumbbell (TD) silicene from molecular calculations: lattice parameters a, b\,(\AA), \comm{and $\gamma$\,($^\circ$)} average cohesive energy $E_{c}$\,(eV/atom), average bond length $d$\,(\AA), average height $h$\,(\AA), 2D elastic constants $\mathrm{C_{ij}}$\,(N/m), 2D Kelvin moduli $K_i$\,(N/m), {total} mean absolute percentage error~(MAPE)\,(\%), {structural mean absolute percentage error~(MAPE$^\textrm{S}$)\,(\%), mechanical mean absolute percentage error~(MAPE$^\textrm{M}$)\,(\%)}. \comm{, relative performance measured as normalized timesteps/second in molecular dynamics (MD) simulation.}}
	\label{tab:Si_TD}
	\centering
	\setlength{\tabcolsep}{1.0pt} 
	\scriptsize
		\begin{tabular}{c c c c c c c c c c}
			\hline
			\textbf{Model} &\textbf{DFT}& \textbf{MEAM}  & \textbf{MACE} & \textbf{MatterSim} & \textbf{CHGNet} & \textbf{Nequix} & \textbf{SevenNet} &\textbf{ORB} &\textbf{PET}\\
			\textbf{version}&           & \textbf{2011}  &\textbf{MP-0b2} &\textbf{v1.0.0-5M} &\textbf{0.3.0} &\textbf{MP-1} &\textbf{L3I5} &\textbf{v3-C.} &\textbf{MAD}\\
							&           &                &\textbf{(large)}&                   &               &              &              & \textbf{OMAT} &             \\
			\hline      a   & 6.434 & 6.511 & 6.518 & 6.556 & 6.648 & 6.623 & 6.588 & 6.537 & 6.613\\
			            b   & 6.434 & 6.511 & 6.518 & 6.556 & 6.648 & 6.623 & 6.588 & 6.537 & 6.613\\
			$-E_{c}$        & 4.679 & 3.865 & 4.077 & 3.346 & 3.662 & 3.699 & 3.919 & 3.114 & 4.437\\
			$d$             & 2.331 & 2.376 & 2.468	& 2.485 & 2.365 & 2.364 & 2.454 & 2.466 & 2.351\\
			$h$             & 2.734 & 2.856 & 2.825 & 2.773 & 2.526 & 2.565 & 2.561 & 2.723 & 2.460\\
			$\mathrm{C_{11}}$ & 100.5 & 80.5 & 72.1 & 69.8 & 80.2\textsuperscript{$\ddag$} & 81.3\textsuperscript{$\ddag$} & 90.8 & 84.5 & 87.7\textsuperscript{$\ddag$}\\
			$\mathrm{C_{22}}$ & 100.5 & 80.5 & 72.1 & 69.8 & 80.1\textsuperscript{$\ddag$} & 81.4\textsuperscript{$\ddag$} & 90.8 & 84.5 & 86.4\textsuperscript{$\ddag$}\\
			$\mathrm{C_{12}}$ & 52.3  & 31.0 & 20.9 & 23.5 & 31.8 & 20.6 & 24.7 & 25.6 & 26.4\\
			$\mathrm{C_{33}}$ & 24.1  & 24.8 & 25.6 & 23.1 & 24.0 & 30.4 & 33.0 & 29.6 & 30.4\\
			$K_{I}$           & 152.8 & 111.5& 93.0 & 93.3 &112.0 & 101.9&115.5 &110.1 &113.5\\
			$K_{II}$          & 48.3  & 49.5 & 51.1 & 46.3 & 48.3\textsuperscript{$\ddag$} &  60.7\textsuperscript{$\ddag$} & 66.1 & 58.9\textsuperscript{$\ddag$} & 60.6\textsuperscript{$\ddag$}\\
			$K_{III}$         & 48.3  & 49.5 & 51.1 & 46.3 & 48.0\textsuperscript{$\ddag$} &  60.8\textsuperscript{$\ddag$} & 66.1 & 59.2\textsuperscript{$\ddag$} & 60.8\textsuperscript{$\ddag$}\\
			MAPE$_\textrm{TD}$              & & 11.809 & 16.522 & 17.307 & 12.071 & 20.367 & 20.000 & 18.423 & 16.760\\
			{MAPE$^\textrm{S}_\textrm{TD}$} & & 2.188  &  2.945 &  2.953 &  3.935 &  3.371 &  4.109 &  2.351 &  4.122\\
			{MAPE$^\textrm{M}_\textrm{TD}$} & & 29.679 & 32.967 & 45.858 & 32.453 & 44.572 & 40.136 & 51.547 & 26.965\\
			\hline
		\end{tabular}
	\begin{tablenotes}[para,flushleft]
		\raggedright
		\small
		\textsuperscript{$\ddag$} {Potential does not reproduce the required isotropy of the elasticity tensor ($\mathrm{C_{11}}\neq\mathrm{C_{22}}$}~\&$/$$\wedge$~$K_{II}$$\neq$$K_{III}$). 
	\end{tablenotes}
\end{table}

\begin{table}[!h]
	\caption{Structural and mechanical properties of honeycomb dumbbell (HD) silicene from molecular calculations: lattice parameters a, b\,(\AA), \comm{and $\gamma$\,($^\circ$)} average cohesive energy $E_{c}$\,(eV/atom), average bond length $d$\,(\AA), average height $h$\,(\AA), 2D elastic constants $\mathrm{C_{ij}}$\,(N/m), 2D Kelvin moduli $K_i$\,(N/m), {total} mean absolute percentage error~(MAPE)\,(\%), {structural mean absolute percentage error~(MAPE$^\textrm{S}$)\,(\%), mechanical mean absolute percentage error~(MAPE$^\textrm{M}$)\,(\%)}. \comm{, relative performance measured as normalized timesteps/second in molecular dynamics (MD) simulation.}}
	\label{tab:Si_HD}
	\centering
	\setlength{\tabcolsep}{1.0pt} 
	\scriptsize
		\begin{tabular}{c c c c c c c c c c}
			\hline
			\textbf{Model} &\textbf{DFT}& \textbf{MEAM}  & \textbf{MACE} & \textbf{MatterSim} & \textbf{CHGNet} & \textbf{Nequix} & \textbf{SevenNet} &\textbf{ORB} &\textbf{PET}\\
			\textbf{version}&           & \textbf{2011}  &\textbf{MP-0b2} &\textbf{v1.0.0-5M} &\textbf{0.3.0} &\textbf{MP-1} &\textbf{L3I5} &\textbf{v3-C.} &\textbf{MAD}\\
							&           &                &\textbf{(large)}&                   &               &              &              & \textbf{OMAT} &             \\
			\hline      a   & 6.297 & 6.364\textsuperscript{$\dag$} & 6.468 & 6.428 & 6.478 & 6.516 & 6.538 & 6.405 & 6.476\\
			            b   & 6.297 & 6.279\textsuperscript{$\dag$} & 6.468 & 6.428 & 6.478 & 6.516 & 6.538 & 6.405 & 6.476\\
			$-E_{c}$        & 4.679 & 4.008 & 4.090 & 3.290 & 3.650 & 3.718 & 3.840 & 3.069 & 4.409\\
			$d$             & 2.399 & 2.405 & 2.402 & 2.431 & 2.403 & 2.422 & 2.397 & 2.408 & 2.418\\
			$h$             & 2.635 & 2.990 & 2.546 & 2.591 & 2.452 & 2.491 & 2.378 & 2.633 & 2.502\\
			$\mathrm{C_{11}}$ & 141.6 & 78.2\textsuperscript{$\ddag$} & 75.8 & 61.3 & 112.8\textsuperscript{$\ddag$} & 65.9 & 77.6 & 84.5 & 71.5\textsuperscript{$\ddag$}\\
			$\mathrm{C_{22}}$ & 141.6 & 82.4\textsuperscript{$\ddag$} & 75.8 & 61.3 & 113.1\textsuperscript{$\ddag$} & 65.9 & 77.6 & 84.5 & 71.3\textsuperscript{$\ddag$}\\
			$\mathrm{C_{12}}$ & 96.4  & 7.1  & 21.5 & 23.4 & 46.0 & 19.3 & 27.0 & 25.6 & 21.2\\
			$\mathrm{C_{33}}$ & 22.6  & 26.2 & 27.1 & 19.0 & 34.3 & 23.3 & 25.3 & 29.6 & 25.3\\
			$K_{I}$           & 238.0 & 87.8 & 97.3 & 84.7 &158.9 & 85.1 &104.5 &110.0 & 92.6\\
			$K_{II}$          & 45.2  & 72.9\textsuperscript{$\ddag$} & 54.3 & 37.9 & 67.0\textsuperscript{$\ddag$} & 46.6 & 50.6 & 58.9\textsuperscript{$\ddag$} & 50.2\textsuperscript{$\ddag$}\\
			$K_{III}$         & 45.2  & 52.3\textsuperscript{$\ddag$} & 54.3 & 37.9 & 68.6\textsuperscript{$\ddag$} & 46.6 & 50.6 & 59.2\textsuperscript{$\ddag$} & 50.5\textsuperscript{$\ddag$}\\
			MAPE$_\textrm{HD}$              & & 30.392 & 25.928 & 28.244 & 25.998 & 24.516 & 24.126 & 28.156 & 24.155\\
			{MAPE$^\textrm{S}_\textrm{HD}$} & & 3.772  &  2.225 &  1.784 &  3.199 &  3.337 &  4.373 &  0.957 &  2.876\\
			{MAPE$^\textrm{M}_\textrm{HD}$} & & 54.466 & 47.221 & 63.743 & 53.897 & 50.521 & 47.451 & 67.559 & 39.134\\
			\hline
		\end{tabular}
	\begin{tablenotes}[para,flushleft]
		\raggedright
		\small
		\textsuperscript{$\dag$} Potential does not reproduce the correct symmetry of the structure (a$\neq$b), \\ 
		\textsuperscript{$\ddag$} {Potential does not reproduce the required isotropy of the elasticity tensor ($\mathrm{C_{11}}\neq\mathrm{C_{22}}$}~\&$/$$\wedge$~$K_{II}$$\neq$$K_{III}$). 
	\end{tablenotes}
\end{table}

\begin{table}[!h]
	\caption{Structural and mechanical properties of large honeycomb dumbbell (LHD) silicene from molecular calculations: lattice parameters a, b\,(\AA), \comm{and $\gamma$\,($^\circ$)} average cohesive energy $E_{c}$\,(eV/atom), average bond length $d$\,(\AA), average height $h$\,(\AA), 2D elastic constants $\mathrm{C_{ij}}$\,(N/m), 2D Kelvin moduli $K_i$\,(N/m), {total} mean absolute percentage error~(MAPE)\,(\%), {structural mean absolute percentage error~(MAPE$^\textrm{S}$)\,(\%), mechanical mean absolute percentage error~(MAPE$^\textrm{M}$)\,(\%)}. \comm{, relative performance measured as normalized timesteps/second in molecular dynamics (MD) simulation.}}
	\label{tab:Si_LHD}
	\centering
	\setlength{\tabcolsep}{1.0pt} 
	\scriptsize
		\begin{tabular}{c c c c c c c c c c}
			\hline
			\textbf{Model} &\textbf{DFT}& \textbf{MEAM}  & \textbf{MACE} & \textbf{MatterSim} & \textbf{CHGNet} & \textbf{Nequix} & \textbf{SevenNet} &\textbf{ORB} &\textbf{PET}\\
			\textbf{version}&           & \textbf{2011}  &\textbf{MP-0b2} &\textbf{v1.0.0-5M} &\textbf{0.3.0} &\textbf{MP-1} &\textbf{L3I5} &\textbf{v3-C.} &\textbf{MAD}\\
							&           &                &\textbf{(large)}&                   &               &              &              & \textbf{OMAT} &             \\
			\hline      a   & 7.334 & 7.363 & 7.474 & 7.408 & 7.556\textsuperscript{$\dag$} & 7.541\textsuperscript{$\dag$} & 7.447 & 7.461 & 7.547\\
			            b   & 7.334 & 7.363 & 7.474 & 7.408 & 7.555\textsuperscript{$\dag$} & 7.540\textsuperscript{$\dag$} & 7.447 & 7.461 & 7.547\\
			$-E_{c}$        & 4.769 & 3.911 & 4.180 & 3.429 & 3.737 & 3.784 & 4.029 & 3.241 & 4.531\\
			$d$             & 2.357 & 2.407 & 2.415 & 2.542 & 2.399 & 2.537 & 2.506 & 2.499 & 2.379\\
			$h$             & 2.683 & 2.857	& 2.678 & 2.825 & 2.578 & 2.800 & 2.638 & 2.656 & 2.457\\
			$\mathrm{C_{11}}$ & 104.5 & 88.0  & 72.1 & 72.6 & 65.9\textsuperscript{$\ddag$} & 72.3 & 97.9 & 84.2 & 80.2\textsuperscript{$\ddag$}\\
			$\mathrm{C_{22}}$ & 104.5 & 88.0  & 72.1 & 72.6 & 66.1\textsuperscript{$\ddag$} & 72.3 & 97.9 & 84.2 & 80.9\textsuperscript{$\ddag$}\\
			$\mathrm{C_{12}}$ & 52.7  & 29.9  & 17.7 & 22.7 & 18.0 & 19.8 & 30.4 & 24.1 & 25.1\\
			$\mathrm{C_{33}}$ & 25.9  & 29.0  & 27.2 & 25.0 & 24.2 & 26.2 & 33.7 & 30.1 & 27.6\\
			$K_{I}$           & 157.2 & 117.9 & 89.7 & 95.3 & 84.0 & 92.0 &128.3 &108.4 &105.6\\
			$K_{II}$          & 51.8  & 58.1  & 54.4 & 49.9 & 48.0\textsuperscript{$\ddag$} & 52.5 & 67.4 & 60.1 & 55.5\textsuperscript{$\ddag$}\\
			$K_{III}$         & 51.8  & 58.1  & 54.4 & 49.9 & 48.5\textsuperscript{$\ddag$} & 52.5 & 67.4 & 60.1 & 55.2\textsuperscript{$\ddag$}\\
			MAPE$_\textrm{LHD}$              & & 13.636 & 17.113 & 17.615 & 19.969 & 17.329 & 15.882 & 17.907 & 14.286\\
			{MAPE$^\textrm{S}_\textrm{LHD}$} & & 2.351  &  1.624 &  3.787 &  2.945 &  4.411 &  2.781 &  2.631 &  3.791\\
			{MAPE$^\textrm{M}_\textrm{LHD}$} & & 32.762 & 34.109 & 45.599 & 44.715 & 39.269 & 34.066 & 49.571 & 23.275\\
			\hline
			$\sum$MAPE$_\textrm{5Si}$             & & 82.553 & 83.762 & 89.123 & 182.977 & 130.171 & 132.332 & 118.370 & 142.379\\
			{$\sum$MAPE$^\textrm{S}_\textrm{5Si}$}& & 22.811 & 35.319 & 17.878 &  46.773 &  38.480 &  27.790 &  13.580 &  36.816\\
			{$\sum$MAPE$^\textrm{M}_\textrm{5Si}$}& &174.224 & 156.150&229.224 & 330.122 & 249.477 & 243.434 & 292.434 & 214.853\\
			\hline
		\end{tabular}
	\begin{tablenotes}[para,flushleft]
		\raggedright
		\small
		\textsuperscript{$\dag$} Potential does not reproduce the correct symmetry of the structure (a$\neq$b), \\ 
		\textsuperscript{$\ddag$} {Potential does not reproduce the required isotropy of the elasticity tensor ($\mathrm{C_{11}}\neq\mathrm{C_{22}}$}~\&$/$$\wedge$~$K_{II}$$\neq$$K_{III}$). 
	\end{tablenotes}
\end{table}

\begin{table}[!h]
	\caption{Structural and mechanical properties of flat (F) germanene from molecular calculations: lattice parameters a, b\,(\AA), \comm{and $\gamma$\,($^\circ$)} average cohesive energy $E_{c}$\,(eV/atom), average bond length $d$\,(\AA), average height $h$\,(\AA), 2D elastic constants $\mathrm{C_{ij}}$\,(N/m), 2D Kelvin moduli $K_i$\,(N/m), {total} mean absolute percentage error~(MAPE)\,(\%), {structural mean absolute percentage error~(MAPE$^\textrm{S}$)\,(\%), mechanical mean absolute percentage error~(MAPE$^\textrm{M}$)\,(\%)}. \comm{, relative performance measured as normalized timesteps/second in molecular dynamics (MD) simulation.}}
	\label{tab:Ge_F}
	\centering
	\setlength{\tabcolsep}{1.0pt} 
	\scriptsize
		\begin{tabular}{c c c c c c c c c c}
			\hline
			\textbf{Model} &\textbf{DFT}& \textbf{SW}    & \textbf{MACE} & \textbf{MatterSim} & \textbf{CHGNet} & \textbf{Nequix} & \textbf{SevenNet} &\textbf{ORB} &\textbf{PET}\\
			\textbf{version}&           & \textbf{1986}  &\textbf{MP-0b2} &\textbf{v1.0.0-5M} &\textbf{0.3.0} &\textbf{MP-1} &\textbf{L3I5} &\textbf{v3-C.} &\textbf{MAD}\\
							&           &                &\textbf{(large)}&                   &               &              &              & \textbf{OMAT} &             \\
			\hline      a   & 4.066 & 4.287 & 4.123 & 4.174 & 4.166 & 4.177 & 4.104 & 4.138 & 4.003\\
			            b   & 4.066 & 4.287 & 4.123 & 4.174 & 4.166 & 4.177 & 4.104 & 4.138 & 4.003\\
			$-E_{c}$        & 3.566 & 2.755 & 3.266 & 2.624 & 2.972 & 3.059 & 3.024 & 2.257 & 4.004\\
			$d$             & 2.349 & 2.475 & 2.384 & 2.412 & 2.405 & 2.412 & 2.370 & 2.391 & 2.328\\
			$h$             & 0.0 & 0.0 & 0.0 & 0.0 & 0.0 & 0.0 & 0.0 & 0.0 & 0.0\\
			$\mathrm{C_{11}}$ & 70.3 & 50.6 & 55.4 & 40.8 & 145.4\textsuperscript{$\ddag$} & 64.4 & 52.8 & 59.0 & 76.3\textsuperscript{$\ddag$}\\
			$\mathrm{C_{22}}$ & 70.3 & 50.6 & 55.4 & 40.8 & 144.9\textsuperscript{$\ddag$} & 64.4 & 52.8 & 59.0 & 76.9\textsuperscript{$\ddag$}\\
			$\mathrm{C_{12}}$ & 29.6 & 25.2 & 18.1 & 12.9 &  18.1 & 23.5 & 10.4 &  7.1 & 18.7\\
			$\mathrm{C_{33}}$ & 20.4 & 12.7 & 18.6 & 14.0 &  64.8 & 20.5 & 21.2 & 25.9 & 28.8\\
			$K_{I}$           & 100.0& 75.8 & 73.5 & 53.7 & 163.3 & 88.0 & 63.2 & 66.1 & 95.2\\
			$K_{II}$          & 40.7 & 25.3 & 37.2 & 27.9 & 127.1\textsuperscript{$\ddag$} & 40.9 & 42.4 & 52.0\textsuperscript{$\ddag$} & 57.9\textsuperscript{$\ddag$}\\
			$K_{III}$         & 40.7 & 25.3 & 37.2 & 27.9 & 129.6\textsuperscript{$\ddag$} & 40.9 & 42.4 & 51.9\textsuperscript{$\ddag$} & 57.6\textsuperscript{$\ddag$}\\
			MAPE$_\textrm{F}$              & & 22.493 &	13.271 & 28.646 & 89.809 & 6.661 & 16.530 & 24.252 & 18.269\\
			{MAPE$^\textrm{S}_\textrm{F}$} & &  5.400 &  1.417 &  2.637 &  2.407 & 2.695 &  0.909 &  1.756 &  1.342\\
			{MAPE$^\textrm{M}_\textrm{F}$} & & 45.949 &	24.027 & 58.209 &135.067 &18.815 & 33.783 & 60.224 & 33.840\\
			\hline
		\end{tabular}
	\begin{tablenotes}[para,flushleft]
		\raggedright
		\small
		\textsuperscript{$\ddag$} {Potential does not reproduce the required isotropy of the elasticity tensor ($\mathrm{C_{11}}\neq\mathrm{C_{22}}$}~\&$/$$\wedge$~$K_{II}$$\neq$$K_{III}$). 
	\end{tablenotes}
\end{table}

\begin{table}[!h]
	\caption{Structural and mechanical properties of low-buckled (LB) germanene from molecular calculations: lattice parameters a, b\,(\AA), \comm{and $\gamma$\,($^\circ$)} average cohesive energy $E_{c}$\,(eV/atom), average bond length $d$\,(\AA), average height $h$\,(\AA), 2D elastic constants $\mathrm{C_{ij}}$\,(N/m), 2D Kelvin moduli $K_i$\,(N/m), {total} mean absolute percentage error~(MAPE)\,(\%), {structural mean absolute percentage error~(MAPE$^\textrm{S}$)\,(\%), mechanical mean absolute percentage error~(MAPE$^\textrm{M}$)\,(\%)}. \comm{, relative performance measured as normalized timesteps/second in molecular dynamics (MD) simulation.}}
	\label{tab:Ge_LB}
	\centering
	\setlength{\tabcolsep}{1.0pt} 
	\scriptsize
		\begin{tabular}{c c c c c c c c c c}
			\hline
			\textbf{Model} &\textbf{DFT}& \textbf{SW}    & \textbf{MACE} & \textbf{MatterSim} & \textbf{CHGNet} & \textbf{Nequix} & \textbf{SevenNet} &\textbf{ORB} &\textbf{PET}\\
			\textbf{version}&           & \textbf{1986}  &\textbf{MP-0b2} &\textbf{v1.0.0-5M} &\textbf{0.3.0} &\textbf{MP-1} &\textbf{L3I5} &\textbf{v3-C.} &\textbf{MAD}\\
							&           &                &\textbf{(large)}&                   &               &              &              & \textbf{OMAT} &             \\
			\hline      a   & 3.955 & 3.998 & 4.045 & 3.948 & 3.615 & 4.177 & 4.075 & 4.044 & 3.989\\
						b   & 3.955 & 3.998 & 4.045 & 3.948 & 3.615 & 4.177 & 4.075 & 4.044 & 3.989\\
			$-E_{c}$        & 3.687 & 2.895 & 3.251 & 2.713 & 3.115 & 3.059 & 3.035 & 2.336 & 4.016\\
			$d$             & 2.370 & 2.447 & 2.426 & 2.433 & 2.524 & 2.412 & 2.391 & 2.440 & 2.328\\
			$h$             & 0.636 & 0.816 & 0.653 & 0.854 & 1.424 & 0.040 & 0.436 & 0.716 & 0.405\\
			$\mathrm{C_{11}}$ & 49.9 & 38.9 & 63.8  & 49.7  & 20.2\textsuperscript{$\ddag$}  & 64.2  & 53.0  & 56.9  & 69.6\textsuperscript{$\ddag$}\\
			$\mathrm{C_{22}}$ & 49.9 & 38.9 & 63.8  & 49.7  & 20.4\textsuperscript{$\ddag$}  & 64.2  & 53.0  & 56.9  & 69.9\textsuperscript{$\ddag$}\\
			$\mathrm{C_{12}}$ & 15.6 &  4.5 & 23.3  & 15.1  &  7.1  & 23.3  & 13.7  &  9.8  & 16.3\\
			$\mathrm{C_{33}}$ & 17.2 & 17.2 & 20.2  & 17.2  &  6.4  & 20.5  & 19.7  & 23.5  & 26.6\\
			$K_{I}$           & 65.4 & 43.4 & 87.1  & 64.8  & 27.4  & 87.5  & 66.7  & 66.7  & 86.1\\
			$K_{II}$          & 34.3 & 34.4 & 40.5  & 34.6\textsuperscript{$\ddag$}  & 13.1\textsuperscript{$\ddag$}  & 40.9  & 39.4\textsuperscript{$\ddag$}  & 47.1  & 53.4\textsuperscript{$\ddag$}\\
			$K_{III}$         & 34.3 & 34.4 & 40.5  & 34.5\textsuperscript{$\ddag$}  & 12.8\textsuperscript{$\ddag$}  & 40.9  & 39.3\textsuperscript{$\ddag$}  & 47.1  & 53.3\textsuperscript{$\ddag$}\\
			MAPE$_\textrm{LB}$              & & 17.067 & 17.842 & 5.831 & 48.419 & 26.904 & 10.583 & 19.646 & 27.607\\
			{MAPE$^\textrm{S}_\textrm{LB}$} & &  8.410 &  2.390 & 9.297 & 36.854 & 26.684 &  9.621 &  5.008 &  9.955\\
			{MAPE$^\textrm{M}_\textrm{LB}$} & & 37.505 & 34.427 &23.903 & 65.842 & 39.796 & 24.321 & 54.441 & 43.135\\
			\hline
		\end{tabular}
	\begin{tablenotes}[para,flushleft]
		\raggedright
		\small
		\textsuperscript{$\ddag$} {Potential does not reproduce the required isotropy of the elasticity tensor ($\mathrm{C_{11}}\neq\mathrm{C_{22}}$}~\&$/$$\wedge$~$K_{II}$$\neq$$K_{III}$). 
	\end{tablenotes}
\end{table}

\begin{table}[!h]
	\caption{Structural and mechanical properties of trigonal dumbbell (TD) germanene from molecular calculations: lattice parameters a, b\,(\AA), \comm{and $\gamma$\,($^\circ$)} average cohesive energy $E_{c}$\,(eV/atom), average bond length $d$\,(\AA), average height $h$\,(\AA), 2D elastic constants $\mathrm{C_{ij}}$\,(N/m), 2D Kelvin moduli $K_i$\,(N/m), {total} mean absolute percentage error~(MAPE)\,(\%), {structural mean absolute percentage error~(MAPE$^\textrm{S}$)\,(\%), mechanical mean absolute percentage error~(MAPE$^\textrm{M}$)\,(\%)}. \comm{, relative performance measured as normalized timesteps/second in molecular dynamics (MD) simulation.}}
	\label{tab:Ge_TD}
	\centering
	\setlength{\tabcolsep}{1.0pt} 
	\scriptsize
		\begin{tabular}{c c c c c c c c c c}
			\hline
			\textbf{Model} &\textbf{DFT}& \textbf{SW}    & \textbf{MACE} & \textbf{MatterSim} & \textbf{CHGNet} & \textbf{Nequix} & \textbf{SevenNet} &\textbf{ORB} &\textbf{PET}\\
			\textbf{version}&           & \textbf{1986}  &\textbf{MP-0b2} &\textbf{v1.0.0-5M} &\textbf{0.3.0} &\textbf{MP-1} &\textbf{L3I5} &\textbf{v3-C.} &\textbf{MAD}\\
							&           &                &\textbf{(large)}&                   &               &              &              & \textbf{OMAT} &             \\
			\hline      a   & 6.698 & 6.933 & 6.878 & 6.982 & 6.949 & 7.040 & 7.006 & 6.905 & 6.861\\
						b   & 6.698 & 6.933 & 6.878 & 6.982 & 6.949 & 7.040 & 7.006 & 6.905 & 6.861\\
			$-E_{c}$        & 3.793 & 2.817 & 3.404 & 2.774 & 3.116 & 3.166 & 3.189 & 2.400 & 4.237\\
			$d$             & 2.449 & 2.596 & 2.655 & 2.533 & 2.516 & 2.510 & 2.502 & 2.520 & 2.458\\
			$h$             & 2.977 & 3.457 & 3.269 & 2.961 & 2.922 & 2.686 & 2.708 & 3.024 & 2.736\\
			$\mathrm{C_{11}}$ & 53.6 & 70.5 & 45.5  & 43.8  & 88.7\textsuperscript{$\ddag$}  & 54.7 &  47.6 & 52.9  & 86.7\textsuperscript{$\ddag$}\\
			$\mathrm{C_{22}}$ & 53.6 & 70.5 & 45.5  & 43.8  & 88.8\textsuperscript{$\ddag$}  & 54.7 &  47.6 & 52.9  & 86.2\textsuperscript{$\ddag$}\\
			$\mathrm{C_{12}}$ & 17.2 & 29.0 & 14.3  & 13.2  & 47.8  & 18.0  &  13.2 & 11.5  & 26.1\\
			$\mathrm{C_{33}}$ & 18.2 & 20.7 & 15.6  & 15.3  & 20.0  & 18.3  &  17.2 & 20.8  & 30.2\\
			$K_{I}$           & 70.8 & 99.5 & 59.7  & 57.0  & 136.6 & 72.7  &  60.8 & 64.4  & 112.5\\
			$K_{II}$          & 36.4 & 41.5 & 31.2  & 30.6  &  40.9\textsuperscript{$\ddag$} & 36.7  &  34.4 & 41.4\textsuperscript{$\ddag$}  &  60.4\\
			$K_{III}$         & 36.4 & 41.5 & 31.2  & 30.6  &  39.9\textsuperscript{$\ddag$} & 36.7  &  34.4 & 41.7\textsuperscript{$\ddag$}  &  60.4\\
			MAPE$_\textrm{TD}$              & & 22.480 & 11.612 & 13.867 & 38.736 & 4.445 & 9.296 & 11.277 & 38.038\\
			{MAPE$^\textrm{S}_\textrm{TD}$} & &  7.287 &  5.892 &  3.115 &  3.020 & 5.625 & 5.096 &  2.662 &  3.334\\
			{MAPE$^\textrm{M}_\textrm{TD}$} & & 49.366 & 22.169 & 39.400 & 69.977 &16.252 & 23.335& 43.129 & 64.169\\
			\hline
		\end{tabular}
	\begin{tablenotes}[para,flushleft]
		\raggedright
		\small
		\textsuperscript{$\ddag$} {Potential does not reproduce the required isotropy of the elasticity tensor ($\mathrm{C_{11}}\neq\mathrm{C_{22}}$}~\&$/$$\wedge$~$K_{II}$$\neq$$K_{III}$). 
	\end{tablenotes}
\end{table}

\begin{table}[!h]
	\caption{Structural and mechanical properties of large honeycomb dumbbell (LHD) germanene from molecular calculations: lattice parameters a, b\,(\AA), \comm{and $\gamma$\,($^\circ$)} average cohesive energy $E_{c}$\,(eV/atom), average bond length $d$\,(\AA), average height $h$\,(\AA), 2D elastic constants $\mathrm{C_{ij}}$\,(N/m), 2D Kelvin moduli $K_i$\,(N/m), {total} mean absolute percentage error~(MAPE)\,(\%), {structural mean absolute percentage error~(MAPE$^\textrm{S}$)\,(\%), mechanical mean absolute percentage error~(MAPE$^\textrm{M}$)\,(\%)}. \comm{, relative performance measured as normalized timesteps/second in molecular dynamics (MD) simulation.}}
	\label{tab:Ge_LHD}
	\centering
	\setlength{\tabcolsep}{1.0pt} 
	\scriptsize
		\begin{tabular}{c c c c c c c c c c}
			\hline
			\textbf{Model} &\textbf{DFT}& \textbf{SW}    & \textbf{MACE} & \textbf{MatterSim} & \textbf{CHGNet} & \textbf{Nequix} & \textbf{SevenNet} &\textbf{ORB} &\textbf{PET}\\
			\textbf{version}&           & \textbf{1986}  &\textbf{MP-0b2} &\textbf{v1.0.0-5M} &\textbf{0.3.0} &\textbf{MP-1} &\textbf{L3I5} &\textbf{v3-C.} &\textbf{MAD}\\
							&           &                &\textbf{(large)}&                   &               &              &              & \textbf{OMAT} &             \\
			\hline      a   & 7.667 & 7.791 & 7.800 & 7.874 & 7.828 & 7.914 & 7.915 & 7.885 & 7.849\\
			            b   & 7.667 & 7.791 & 7.800 & 7.874 & 7.828 & 7.914 & 7.915 & 7.885 & 7.849\\
			$-E_{c}$        & 3.863 & 2.844 & 3.448 & 2.859 & 3.214 & 3.246 & 3.297 & 2.501 & 4.329\\
			$d$             & 2.485 & 2.642 & 2.589 & 2.569 & 2.542 & 2.537 & 2.542 & 2.542 & 2.498\\
			$h$             & 2.875 & 3.449 & 3.209 & 3.025 & 2.950 & 2.800 & 2.868 & 2.880 & 2.715\\
			$\mathrm{C_{11}}$ & 57.7& 76.3  & 39.4  & 47.0  &  42.2\textsuperscript{$\ddag$} &  53.9 &  37.8 & 42.7  & 58.3\textsuperscript{$\ddag$}\\
			$\mathrm{C_{22}}$ & 57.7& 76.3  & 39.4  & 47.0  &  42.3\textsuperscript{$\ddag$} &  53.9 &  37.8 & 42.7  & 58.6\textsuperscript{$\ddag$}\\
			$\mathrm{C_{12}}$ & 20.7& 35.4  & 12.5  & 12.2  &  11.0 &  17.8 &  7.2  & 11.2  & 14.7\\
			$\mathrm{C_{33}}$ & 18.5& 20.4  & 13.4  & 17.4  &  15.7 &  18.1 &  15.3 & 15.8  & 21.9\\
			$K_{I}$           & 78.4& 111.7 & 51.9  & 59.2  &  53.3 &  71.7 &  45.1 & 53.9  & 73.2\\
			$K_{II}$          & 37.0& 40.9  & 26.9  & 34.8  &  31.2\textsuperscript{$\ddag$} &  36.2 &  30.6 & 31.5\textsuperscript{$\ddag$}  & 43.8\textsuperscript{$\ddag$}\\
			$K_{III}$         & 37.0& 40.9  & 26.9  & 34.8  &  31.4\textsuperscript{$\ddag$} &  36.2 &  30.6 & 31.7\textsuperscript{$\ddag$}  & 43.7\textsuperscript{$\ddag$}\\
			MAPE$_\textrm{LHD}$              & & 22.085 & 20.727 & 13.416 & 17.027 & 5.845 & 21.011 & 18.031 & 9.634\\
			{MAPE$^\textrm{S}_\textrm{LHD}$} & & 7.378  &  4.811 &  3.502 &  2.273 & 2.803 &  2.260 &  2.044 & 2.714\\
			{MAPE$^\textrm{M}_\textrm{LHD}$} & & 49.210 & 36.731 & 37.869 & 37.008 &19.342 & 41.361 & 52.467 &22.145\\
			\hline
			$\sum$MAPE$_\textrm{4Ge}$             & &84.125  & 63.451 & 61.761 & 193.991 & 43.854 & 57.419 & 73.206 & 93.547\\
			{$\sum$MAPE$^\textrm{S}_\textrm{4Ge}$}& & 28.475 & 14.510 & 18.551 &  44.555 & 37.806 & 17.885 & 11.469 & 17.346\\
			{$\sum$MAPE$^\textrm{M}_\textrm{4Ge}$}& &182.030 & 117.354&159.380 & 307.893 & 94.205 &122.800 &210.261 &163.289\\
			\hline
		\end{tabular}
	\begin{tablenotes}[para,flushleft]
		\raggedright
		\small
		\textsuperscript{$\ddag$} {Potential does not reproduce the required isotropy of the elasticity tensor ($\mathrm{C_{11}}\neq\mathrm{C_{22}}$}~\&$/$$\wedge$~$K_{II}$$\neq$$K_{III}$). 
	\end{tablenotes}
\end{table}

\begin{table}[!h]
	\caption{Structural and mechanical properties of flat (F) stanene from molecular calculations: lattice parameters a, b\,(\AA), \comm{and $\gamma$\,($^\circ$)} average cohesive energy $E_{c}$\,(eV/atom), average bond length $d$\,(\AA), average height $h$\,(\AA), 2D elastic constants $\mathrm{C_{ij}}$\,(N/m), 2D Kelvin moduli $K_i$\,(N/m), {total} mean absolute percentage error~(MAPE)\,(\%), {structural mean absolute percentage error~(MAPE$^\textrm{S}$)\,(\%), mechanical mean absolute percentage error~(MAPE$^\textrm{M}$)\,(\%)}. \comm{, relative performance measured as normalized timesteps/second in molecular dynamics (MD) simulation.}}
	\label{tab:Sn_F}
	\centering
	\setlength{\tabcolsep}{1.0pt} 
	\scriptsize
		\begin{tabular}{c c c c c c c c c c}
			\hline
			\textbf{Model} &\textbf{DFT}& \textbf{MEAM}  & \textbf{MACE} & \textbf{MatterSim} & \textbf{CHGNet} & \textbf{Nequix} & \textbf{SevenNet} &\textbf{ORB} &\textbf{PET}\\
			\textbf{version}&           & \textbf{2018b} &\textbf{MP-0b2} &\textbf{v1.0.0-5M} &\textbf{0.3.0} &\textbf{MP-1} &\textbf{L3I5} &\textbf{v3-C.} &\textbf{MAD}\\
							&           &                &\textbf{(large)}&                   &               &              &              & \textbf{OMAT} &             \\
			\hline      a   & 4.753 & 4.939 & 4.753 & 4.936 & 4.884 & 4.955 & 4.904 & 4.814 & 4.833\\
						b   & 4.753 & 4.939 & 4.753 & 4.936 & 4.884 & 4.955 & 4.904 & 4.814 & 4.833\\
			$-E_{c}$        & 2.719 & 2.656 & 2.524 & 2.009 & 2.091 & 2.570 & 2.482 & 1.726 & 2.572\\
			$d$             & 2.746 & 2.851 & 2.748 & 2.853 & 2.818 & 2.860 & 2.832 & 2.783 & 2.783\\
			$h$             & 0.0 & 0.0 & 0.0 & 0.0 & 0.0 & 0.0 & 0.0 & 0.0 & 0.0\\
			$\mathrm{C_{11}}$ & 31.3 & 27.1 & 19.2 & 22.5 & 13.7\textsuperscript{$\ddag$} & 28.3 & 17.4 & 45.5\textsuperscript{$\ddag$} & 23.8\textsuperscript{$\ddag$}\\
			$\mathrm{C_{22}}$ & 31.3 & 27.1 & 19.2 & 22.5 & 13.8\textsuperscript{$\ddag$} & 28.3 & 17.4 & 45.4\textsuperscript{$\ddag$} & 24.0\textsuperscript{$\ddag$}\\
			$\mathrm{C_{12}}$ & 17.7 & 19.8 &  8.2 &  6.7 &  2.9 & 11.2 &  1.6 & 14.6 &  3.7\\
			$\mathrm{C_{33}}$ &  6.8 &  3.7 &  5.5 &  7.9 &  5.4 &  8.6 &  7.9 & 15.4 & 10.1\\
			$K_{I}$           & 49.0 & 46.8 & 27.4 & 29.2 & 16.7 & 39.5 & 19.0 & 60.1 & 27.6\\
			$K_{II}$          & 13.6 &  7.3 & 11.0 & 15.8 & 10.8 & 17.2 & 15.8\textsuperscript{$\ddag$} & 30.8 & 20.2\textsuperscript{$\ddag$}\\
			$K_{III}$         & 13.6 &  7.3 & 11.0 & 15.8 & 10.8 & 17.2 & 15.7\textsuperscript{$\ddag$} & 30.8 & 20.3\textsuperscript{$\ddag$}\\
			MAPE$_\textrm{F}$              & & 17.831 & 21.803 & 22.322 & 32.268 & 15.652 & 27.801 & 49.837 & 29.617\\
			{MAPE$^\textrm{S}_\textrm{F}$} & &  2.910 &  0.022 &  3.856 &  2.707 &  4.212 &  3.147 &  1.299 &  1.565\\
			{MAPE$^\textrm{M}_\textrm{F}$} & & 24.804 & 35.360 & 48.842 & 60.686 & 24.064 & 43.571 & 95.434 & 44.177\\
			\hline
		\end{tabular}
	\begin{tablenotes}[para,flushleft]
		\raggedright
		\small
		\textsuperscript{$\ddag$} {Potential does not reproduce the required isotropy of the elasticity tensor ($\mathrm{C_{11}}\neq\mathrm{C_{22}}$}~\&$/$$\wedge$~$K_{II}$$\neq$$K_{III}$). 
	\end{tablenotes}
\end{table}

\begin{table}[!h]
	\caption{Structural and mechanical properties of low-buckled (LB) stanene from molecular calculations: lattice parameters a, b\,(\AA), \comm{and $\gamma$\,($^\circ$)} average cohesive energy $E_{c}$\,(eV/atom), average bond length $d$\,(\AA), average height $h$\,(\AA), 2D elastic constants $\mathrm{C_{ij}}$\,(N/m), 2D Kelvin moduli $K_i$\,(N/m), {total} mean absolute percentage error~(MAPE)\,(\%), {structural mean absolute percentage error~(MAPE$^\textrm{S}$)\,(\%), mechanical mean absolute percentage error~(MAPE$^\textrm{M}$)\,(\%)}. \comm{, relative performance measured as normalized timesteps/second in molecular dynamics (MD) simulation.}}
	\label{tab:Sn_LB}
	\centering
	\setlength{\tabcolsep}{1.0pt} 
	\scriptsize
		\begin{tabular}{c c c c c c c c c c}
			\hline
			\textbf{Model} &\textbf{DFT}& \textbf{MEAM}  & \textbf{MACE} & \textbf{MatterSim} & \textbf{CHGNet} & \textbf{Nequix} & \textbf{SevenNet} &\textbf{ORB} &\textbf{PET}\\
			\textbf{version}&           & \textbf{2018b} &\textbf{MP-0b2} &\textbf{v1.0.0-5M} &\textbf{0.3.0} &\textbf{MP-1} &\textbf{L3I5} &\textbf{v3-C.} &\textbf{MAD}\\
							&           &                &\textbf{(large)}&                   &               &              &              & \textbf{OMAT} &             \\
			\hline      a   & 4.594 & 4.500 & 4.045 & 4.243 & 3.306\textsuperscript{$\dag$} & 4.456 & 4.536 & 4.621\textsuperscript{$\dag$} & 4.462\\
						b   & 4.594 & 4.500 & 4.045 & 4.243 & 3.311\textsuperscript{$\dag$} & 4.456 & 4.536 & 4.620\textsuperscript{$\dag$} & 4.462\\
			$-E_{c}$        & 2.844 & 2.743 & 2.856 & 2.134 & 2.482 & 2.710 & 2.734 & 1.800 & 2.697\\
			$d$             & 2.791 & 2.823 & 2.895 & 2.923 & 3.304 & 2.867 & 2.916 & 2.832 & 2.799\\
			$h$             & 0.875 & 1.104 & 1.484 & 1.595 & 2.685 & 1.268 & 1.285 & 0.952 & 1.093\\
			$\mathrm{C_{11}}$ & 23.4 & 11.7 & 29.2 & 29.0 & 39.0\textsuperscript{$\ddag$} & 33.1 & 35.2 & 41.2\textsuperscript{$\ddag$} & 171.0\textsuperscript{$\ddag$}\\
			$\mathrm{C_{22}}$ & 23.4 & 11.7 & 29.2 & 29.0 & 34.4\textsuperscript{$\ddag$} & 33.1 & 35.2 & 41.0\textsuperscript{$\ddag$} & 167.3\textsuperscript{$\ddag$}\\
			$\mathrm{C_{12}}$ &  8.1 &  3.1 & 11.6 & 15.9 & 30.5 & 13.8 & 12.2 &  9.5 &  53.8\\
			$\mathrm{C_{33}}$ &  7.7 &  4.3 &  8.8 &  6.6 &  2.0 &  9.7 & 11.5 & 15.8 &  55.7\\
			$K_{I}$           & 31.5 & 14.8 & 40.8 & 44.9 & 67.1 & 47.0 & 47.4 & 50.6 & 223.0\\
			$K_{II}$          & 15.3 &  8.6 & 17.6 & 13.2\textsuperscript{$\ddag$} &  6.3\textsuperscript{$\ddag$} & 19.3 & 23.0 & 31.6\textsuperscript{$\ddag$} & 115.4\textsuperscript{$\ddag$}\\
			$K_{III}$         & 15.3 &  8.6 & 17.6 & 13.1\textsuperscript{$\ddag$} &  4.0\textsuperscript{$\ddag$} & 19.3 & 23.0 & 31.7\textsuperscript{$\ddag$} & 111.4\textsuperscript{$\ddag$}\\
			MAPE$_\textrm{LB}$              & & 31.725 & 21.971 & 29.771 & 83.701 & 28.270 & 34.171 & 49.660 & 363.151\\
			{MAPE$^\textrm{S}_\textrm{LB}$} & &  7.854 & 24.316 & 25.583 & 70.303 & 13.416 & 13.474 &  2.844 &   7.735\\
			{MAPE$^\textrm{M}_\textrm{LB}$} & & 46.311 & 21.133 & 50.574 & 99.946 & 39.229 & 47.408 &100.593 & 544.727\\
			\hline
		\end{tabular}
	\begin{tablenotes}[para,flushleft]
		\raggedright
		\small
		\textsuperscript{$\dag$} Potential does not reproduce the correct symmetry of the structure (a$\neq$b), \\ 
		\textsuperscript{$\ddag$} {Potential does not reproduce the required isotropy of the elasticity tensor ($\mathrm{C_{11}}\neq\mathrm{C_{22}}$}~\&$/$$\wedge$~$K_{II}$$\neq$$K_{III}$). 
	\end{tablenotes}
\end{table}

\begin{table}[!h]
	\caption{Structural and mechanical properties of high-buckled (HB) stanene from molecular calculations: lattice parameters a, b\,(\AA), \comm{and $\gamma$\,($^\circ$)} average cohesive energy $E_{c}$\,(eV/atom), average bond length $d$\,(\AA), average height $h$\,(\AA), 2D elastic constants $\mathrm{C_{ij}}$\,(N/m), 2D Kelvin moduli $K_i$\,(N/m), {total} mean absolute percentage error~(MAPE)\,(\%), {structural mean absolute percentage error~(MAPE$^\textrm{S}$)\,(\%), mechanical mean absolute percentage error~(MAPE$^\textrm{M}$)\,(\%)}. \comm{, relative performance measured as normalized timesteps/second in molecular dynamics (MD) simulation.}}
	\label{tab:Sn_HB}
	\centering
	\setlength{\tabcolsep}{1.0pt} 
	\scriptsize
		\begin{tabular}{c c c c c c c c c c}
			\hline
			\textbf{Model} &\textbf{DFT}& \textbf{MEAM}  & \textbf{MACE} & \textbf{MatterSim} & \textbf{CHGNet} & \textbf{Nequix} & \textbf{SevenNet} &\textbf{ORB} &\textbf{PET}\\
			\textbf{version}&           & \textbf{2018b} &\textbf{MP-0b2} &\textbf{v1.0.0-5M} &\textbf{0.3.0} &\textbf{MP-1} &\textbf{L3I5} &\textbf{v3-C.} &\textbf{MAD}\\
							&           &                &\textbf{(large)}&                   &               &              &              & \textbf{OMAT} &             \\
			\hline      a   & 3.329 & 3.232 & 3.501 & 3.319 & 3.319 & 3.309 & 3.330 & 3.327\textsuperscript{$\dag$} & 3.254\\
						b   & 3.329 & 3.232 & 3.501 & 3.319 & 3.319 & 3.309 & 3.330 & 3.330\textsuperscript{$\dag$} & 3.254\\
			$-E_{c}$        & 3.179 & 2.885 & 2.965 & 2.382 & 2.482 & 2.857 & 2.887 & 2.036 & 3.165\\
			$d$             & 3.266 & 3.217 & 3.133 & 3.302 & 3.310 & 3.274 & 3.324 & 3.252 & 3.244 \\
			$h$             & 2.497 & 2.585 & 2.394 & 2.647 & 2.670 & 2.581 & 2.699 & 2.628 & 2.626\\
			$\mathrm{C_{11}}$ & 44.0 & 89.4 & 30.5 & 49.7 & 25.5 & 38.5 & 33.9 & 62.9\textsuperscript{$\ddag$} & 47.5\textsuperscript{$\ddag$}\\
			$\mathrm{C_{22}}$ & 44.0 & 89.4 & 30.5 & 49.7 & 25.5 & 38.5 & 33.9 & 65.1\textsuperscript{$\ddag$} & 46.7\textsuperscript{$\ddag$}\\
			$\mathrm{C_{12}}$ & 28.6 & 28.8 & 20.5 & 33.1 & 26.9 & 27.1 & 30.4 & 45.3 & 22.4\\
			$\mathrm{C_{33}}$ &  7.7 & 30.3 &  5.0 &  8.3 & -0.8 &  5.7 &  1.8 &  8.2 & 12.2\\
			$K_{I}$           & 72.7 &118.3 & 50.9 & 82.7 & 52.4 & 65.6 & 64.3 &109.2 & 69.5\\
			$K_{II}$          & 15.4 & 60.6 & 10.0 & 16.6 & -1.4\textsuperscript{$\ddag$}\textsuperscript{$\ast$} & 11.4 &  3.6\textsuperscript{$\ddag$} & 18.7\textsuperscript{$\ddag$} & 24.7\textsuperscript{$\ddag$}\\
			$K_{III}$         & 15.4 & 60.6 & 10.0 & 16.6 & -1.6\textsuperscript{$\ddag$}\textsuperscript{$\ast$} & 11.4 &  3.5\textsuperscript{$\ddag$} & 16.5\textsuperscript{$\ddag$} & 24.5\textsuperscript{$\ddag$}\\
			MAPE$_\textrm{HB}$              & & 97.449 & 20.902 & 9.199 & 39.889 & 11.145 & 26.124 & 22.963 & 19.051\\
			{MAPE$^\textrm{S}_\textrm{HB}$} & &  2.718 &  4.631 & 1.938 &  2.228 &  1.189 &  2.488 &  1.444 &  2.592\\
			{MAPE$^\textrm{M}_\textrm{HB}$} & & 151.752& 34.089 &31.636 & 75.173 & 23.731 & 44.840 & 60.700 & 27.624\\
			\hline
		\end{tabular}
	\begin{tablenotes}[para,flushleft]
		\raggedright
		\small
		\textsuperscript{$\dag$} Potential does not reproduce the correct symmetry of the structure (a$\neq$b), \\ 
		\textsuperscript{$\ast$} Negative Kelvin moduli $K_i$ indicating a lack of mechanical stability,\\
		\textsuperscript{$\ddag$} {Potential does not reproduce the required isotropy of the elasticity tensor ($\mathrm{C_{11}}\neq\mathrm{C_{22}}$}~\&$/$$\wedge$~$K_{II}$$\neq$$K_{III}$). 
	\end{tablenotes}
\end{table}

\begin{table}[!h]
	\caption{Structural and mechanical properties of full dumbbell (FD) stanene from molecular calculations: lattice parameters a, b\,(\AA), \comm{and $\gamma$\,($^\circ$)} average cohesive energy $E_{c}$\,(eV/atom), average bond length $d$\,(\AA), average height $h$\,(\AA), 2D elastic constants $\mathrm{C_{ij}}$\,(N/m), 2D Kelvin moduli $K_i$\,(N/m), {total} mean absolute percentage error~(MAPE)\,(\%), {structural mean absolute percentage error~(MAPE$^\textrm{S}$)\,(\%), mechanical mean absolute percentage error~(MAPE$^\textrm{M}$)\,(\%)}. \comm{, relative performance measured as normalized timesteps/second in molecular dynamics (MD) simulation.}}
	\label{tab:Sn_FD}
	\centering
	\setlength{\tabcolsep}{1.0pt} 
	\scriptsize
		\begin{tabular}{c c c c c c c c c c}
			\hline
			\textbf{Model} &\textbf{DFT}& \textbf{MEAM}  & \textbf{MACE} & \textbf{MatterSim} & \textbf{CHGNet} & \textbf{Nequix} & \textbf{SevenNet} &\textbf{ORB} &\textbf{PET}\\
			\textbf{version}&           & \textbf{2018b} &\textbf{MP-0b2} &\textbf{v1.0.0-5M} &\textbf{0.3.0} &\textbf{MP-1} &\textbf{L3I5} &\textbf{v3-C.} &\textbf{MAD}\\
							&           &                &\textbf{(large)}&                   &               &              &              & \textbf{OMAT} &             \\
			\hline      a   & 4.309 & 4.359 & 4.469 & 4.462 & 4.516 & 4.498 & 4.566 & 4.378\textsuperscript{$\dag$} & 4.268\\
						b   & 4.309 & 4.359 & 4.469 & 4.462 & 4.516 & 4.498 & 4.566 & 4.377\textsuperscript{$\dag$} & 4.268\\
			$-E_{c}$        & 3.043 & 2.782 & 2.973 & 2.257 & 2.361 & 2.811 & 2.924 & 1.996 & 2.977\\
			$d$             & 3.012 & 2.998 & 2.999 & 3.028 & 3.034 & 3.019 & 3.028 & 3.038 & 2.961\\
			$h$             & 3.267 & 3.170 & 3.045 & 3.118 & 3.078 & 3.056 & 2.994 & 3.260 & 3.177\\
			$\mathrm{C_{11}}$ & 15.3 & 24.4 & 33.1 & 35.9 & 22.6 & 30.3 & 40.7 & 27.7 & 41.8\textsuperscript{$\ddag$}\\
			$\mathrm{C_{22}}$ & 15.3 & 24.4 & 33.1 & 35.9 & 22.6 & 30.3 & 40.7 & 28.0 & 42.1\textsuperscript{$\ddag$}\\
			$\mathrm{C_{12}}$ &  8.7 & 12.9 &  8.7 & 15.2 & 13.4 & 17.8 & 17.3 & 15.6 & 21.5\\
			$\mathrm{C_{33}}$ &  3.3 &  5.8 & 12.2 & 10.3 &  4.6 &  6.2 & 11.7 &  6.0 & 10.1\\
			$K_{I}$           & 24.0 & 37.3 & 41.8 & 51.1 & 36.0 & 48.1 & 58.0 & 43.5 & 63.4\\
			$K_{II}$          &  6.5 & 11.5 & 24.4 & 20.6 &  9.1 & 12.5 & 23.5 & 12.3\textsuperscript{$\ddag$} & 20.5\textsuperscript{$\ddag$}\\
			$K_{III}$         &  6.5 & 11.5 & 24.4 & 20.6 &  9.1 & 12.5 & 23.5 & 11.9\textsuperscript{$\ddag$} & 20.3\textsuperscript{$\ddag$}\\
			MAPE$_\textrm{FD}$              & & 42.292 & 104.132 & 103.919 & 32.522 & 63.341 & 125.077 & 56.180 & 118.488\\
			{MAPE$^\textrm{S}_\textrm{FD}$} & &  1.436 &   3.663 &   3.050 &  4.028 &  3.866 &   5.205 &  1.065 &   1.585\\
			{MAPE$^\textrm{M}_\textrm{FD}$} & & 63.858 & 143.068 & 160.720 & 59.513 & 90.868 & 172.309 &102.511 & 163.743\\
			\hline
		\end{tabular}
	\begin{tablenotes}[para,flushleft]
		\raggedright
		\small
		\textsuperscript{$\dag$} Potential does not reproduce the correct symmetry of the structure (a$\neq$b), \\ 
		\textsuperscript{$\ddag$} {Potential does not reproduce the required isotropy of the elasticity tensor ($\mathrm{C_{11}}\neq\mathrm{C_{22}}$}~\&$/$$\wedge$~$K_{II}$$\neq$$K_{III}$). 
	\end{tablenotes}
\end{table}

\begin{table}[!h]
	\caption{Structural and mechanical properties of trigonal dumbbell (TD) stanene from molecular calculations: lattice parameters a, b\,(\AA), \comm{and $\gamma$\,($^\circ$)} average cohesive energy $E_{c}$\,(eV/atom), average bond length $d$\,(\AA), average height $h$\,(\AA), 2D elastic constants $\mathrm{C_{ij}}$\,(N/m), 2D Kelvin moduli $K_i$\,(N/m), {total} mean absolute percentage error~(MAPE)\,(\%), {structural mean absolute percentage error~(MAPE$^\textrm{S}$)\,(\%), mechanical mean absolute percentage error~(MAPE$^\textrm{M}$)\,(\%)}. \comm{, relative performance measured as normalized timesteps/second in molecular dynamics (MD) simulation.}}
	\label{tab:Sn_TD}
	\centering
	\setlength{\tabcolsep}{1.0pt} 
	\scriptsize
		\begin{tabular}{c c c c c c c c c c}
			\hline
			\textbf{Model} &\textbf{DFT}& \textbf{MEAM}  & \textbf{MACE} & \textbf{MatterSim} & \textbf{CHGNet} & \textbf{Nequix} & \textbf{SevenNet} &\textbf{ORB} &\textbf{PET}\\
			\textbf{version}&           & \textbf{2018b} &\textbf{MP-0b2} &\textbf{v1.0.0-5M} &\textbf{0.3.0} &\textbf{MP-1} &\textbf{L3I5} &\textbf{v3-C.} &\textbf{MAD}\\
							&           &                &\textbf{(large)}&                   &               &              &              & \textbf{OMAT} &             \\
			\hline      a   & 7.821 & 7.951 & 8.004 & 8.226 & 8.241 & 8.208 & 8.187 & 8.016\textsuperscript{$\dag$} & 7.961\\
						b   & 7.821 & 7.951 & 8.004 & 8.226 & 8.241 & 8.208 & 8.187 & 8.014\textsuperscript{$\dag$} & 7.961\\
			$-E_{c}$        & 2.933 & 2.773 & 2.800 & 2.165 & 2.236 & 2.720 & 2.763 & 1.889 & 2.929\\
			$d$             & 2.830 & 2.885 & 2.901 & 2.943 & 2.931 & 2.936 & 2.939 & 2.907 & 2.836\\
			$h$             & 3.514 & 3.638 & 3.620 & 3.228 & 3.133 & 3.220 & 3.289 & 3.408 & 3.242\\
			$\mathrm{C_{11}}$ & 27.7&  17.1 & 29.1 & 30.7 & 41.5 & 30.6 & 26.1 & 40.2\textsuperscript{$\ddag$} & 87.5\textsuperscript{$\ddag$}\\
			$\mathrm{C_{22}}$ & 27.7&  17.1 & 29.1 & 30.7 & 41.5 & 30.6 & 26.1 & 41.0\textsuperscript{$\ddag$} & 86.2\textsuperscript{$\ddag$}\\
			$\mathrm{C_{12}}$ & 11.0&   8.4 & 10.1 & 11.4 & 25.7 & 12.1 &  5.9 & 11.9 & 26.4\\
			$\mathrm{C_{33}}$ &  8.3&   4.4 &  9.5 &  9.7 &  7.9 &  9.3 & 10.1 & 14.5 & 30.3\\
			$K_{I}$           & 38.7&  25.4 & 39.2 & 42.1 & 67.2 & 42.6 & 32.0 & 52.5 & 113.2\\
			$K_{II}$          & 16.6&   8.7 & 19.0 & 19.3 & 15.8 & 18.5 & 20.2 & 28.7\textsuperscript{$\ddag$} &  60.5\textsuperscript{$\ddag$}\\
			$K_{III}$         & 16.6&   8.7 & 19.0 & 19.3 & 15.8 & 18.5 & 20.2 & 29.1\textsuperscript{$\ddag$} &  60.6\textsuperscript{$\ddag$}\\
			MAPE$_\textrm{TD}$              & & 24.294 & 6.496 & 11.014 & 30.836 & 8.711 & 13.730 & 33.868 & 130.487\\
			{MAPE$^\textrm{S}_\textrm{TD}$} & &  2.195 & 2.551 &  5.618 &  6.281 & 5.492 &  4.896 &  2.666 &   2.878\\
			{MAPE$^\textrm{M}_\textrm{TD}$} & & 39.422 &11.853 & 33.341 & 60.919 & 15.771& 22.488 & 76.162 & 194.396\\
			\hline
		\end{tabular}
	\begin{tablenotes}[para,flushleft]
		\raggedright
		\small
		\textsuperscript{$\dag$} Potential does not reproduce the correct symmetry of the structure (a$\neq$b), \\ 
		\textsuperscript{$\ddag$} {Potential does not reproduce the required isotropy of the elasticity tensor ($\mathrm{C_{11}}\neq\mathrm{C_{22}}$}~\&$/$$\wedge$~$K_{II}$$\neq$$K_{III}$). 
	\end{tablenotes}
\end{table}

\begin{table}[!h]
	\caption{Structural and mechanical properties of honeycomb dumbbell (HD) stanene from molecular calculations: lattice parameters a, b\,(\AA), \comm{and $\gamma$\,($^\circ$)} average cohesive energy $E_{c}$\,(eV/atom), average bond length $d$\,(\AA), average height $h$\,(\AA), 2D elastic constants $\mathrm{C_{ij}}$\,(N/m), 2D Kelvin moduli $K_i$\,(N/m), {total} mean absolute percentage error~(MAPE)\,(\%), {structural mean absolute percentage error~(MAPE$^\textrm{S}$)\,(\%), mechanical mean absolute percentage error~(MAPE$^\textrm{M}$)\,(\%)}. \comm{, relative performance measured as normalized timesteps/second in molecular dynamics (MD) simulation.}}
	\label{tab:Sn_HD}
	\centering
	\setlength{\tabcolsep}{1.0pt} 
	\scriptsize
		\begin{tabular}{c c c c c c c c c c}
			\hline
			\textbf{Model} &\textbf{DFT}& \textbf{MEAM}  & \textbf{MACE} & \textbf{MatterSim} & \textbf{CHGNet} & \textbf{Nequix} & \textbf{SevenNet} &\textbf{ORB} &\textbf{PET}\\
			\textbf{version}&           & \textbf{2018b} &\textbf{MP-0b2} &\textbf{v1.0.0-5M} &\textbf{0.3.0} &\textbf{MP-1} &\textbf{L3I5} &\textbf{v3-C.} &\textbf{MAD}\\
							&           &                &\textbf{(large)}&                   &               &              &              & \textbf{OMAT} &             \\
			\hline      a   & 7.646 & 7.730 & 7.905 & 8.010 & 8.013 & 7.990 & 8.055 & 7.778 & 7.608\\
						b   & 7.646 & 7.730 & 7.905 & 8.010 & 8.013 & 7.990 & 8.055 & 7.778 & 7.608\\
			$-E_{c}$        & 3.040 & 2.788 & 2.917 & 2.223 & 2.313 & 2.787 & 2.878 & 1.947 & 2.919\\
			$d$             & 2.945 & 2.966 & 2.965 & 2.988 & 2.995 & 2.981 & 2.988 & 2.992 & 2.907\\
			$h$             & 3.290 & 3.288 & 3.101 & 3.096 & 3.114 & 3.090 & 3.044 & 3.337 & 3.189\\
			$\mathrm{C_{11}}$ & 27.9& 17.8  & 31.9 & 31.7 & 38.2 & 33.4 & 35.9 & 42.0\textsuperscript{$\ddag$} & 71.6\textsuperscript{$\ddag$}\\
			$\mathrm{C_{22}}$ & 27.9& 17.8  & 31.9 & 31.7 & 38.2 & 33.4 & 35.9 & 41.8\textsuperscript{$\ddag$} & 71.3\textsuperscript{$\ddag$}\\
			$\mathrm{C_{12}}$ & 13.1&  7.3  &  9.6 & 12.9 & 24.4 & 14.9 & 12.4 & 15.0 & 21.2\\
			$\mathrm{C_{33}}$ &  7.4&  5.2  & 11.2 &  9.4 &  6.9 &  9.2 & 11.8 & 13.4 & 25.3\\
			$K_{I}$           & 41.0& 25.1  & 41.5 & 44.6 & 62.6 & 48.3 & 48.3 & 56.9 & 92.6\\
			$K_{II}$          & 14.8& 10.5  & 22.3 & 18.8 & 13.9 & 18.4 & 23.5 & 26.9\textsuperscript{$\ddag$} & 50.3\textsuperscript{$\ddag$}\\
			$K_{III}$         & 14.8& 10.5  & 22.3 & 18.8 & 13.9 & 18.4 & 23.5 & 26.7\textsuperscript{$\ddag$} & 50.5\textsuperscript{$\ddag$}\\
			MAPE$_\textrm{HD}$              & & 21.219 & 18.696 & 13.471 & 22.753 & 14.024 & 23.361 & 36.450 & 102.376\\
			{MAPE$^\textrm{S}_\textrm{HD}$} & &  0.742 &  3.305 &  4.232 &  4.171 &  4.082 &  4.918 &  1.618 &   1.339\\
			{MAPE$^\textrm{M}_\textrm{HD}$} & & 37.679 & 29.411 & 38.234 & 49.982 & 25.235 & 36.578 & 80.835 & 155.869\\
			\hline
		\end{tabular}
	\begin{tablenotes}[para,flushleft]
		\raggedright
		\small
		\textsuperscript{$\ddag$} {Potential does not reproduce the required isotropy of the elasticity tensor ($\mathrm{C_{11}}\neq\mathrm{C_{22}}$}~\&$/$$\wedge$~$K_{II}$$\neq$$K_{III}$). 
	\end{tablenotes}
\end{table}

\begin{table}[!h]
	\caption{Structural and mechanical properties of large honeycomb dumbbell (LHD) stanene from molecular calculations: lattice parameters a, b\,(\AA), \comm{and $\gamma$\,($^\circ$)} average cohesive energy $E_{c}$\,(eV/atom), average bond length $d$\,(\AA), average height $h$\,(\AA), 2D elastic constants $\mathrm{C_{ij}}$\,(N/m), 2D Kelvin moduli $K_i$\,(N/m), {total} mean absolute percentage error~(MAPE)\,(\%), {structural mean absolute percentage error~(MAPE$^\textrm{S}$)\,(\%), mechanical mean absolute percentage error~(MAPE$^\textrm{M}$)\,(\%)}, standard deviation $\sigma$.\comm{, relative performance measured as normalized timesteps/second in molecular dynamics (MD) simulation.}}
	\label{tab:Sn_LHD}
	\centering
	\setlength{\tabcolsep}{1.0pt} 
	\scriptsize
		\begin{tabular}{c c c c c c c c c c}
			\hline
			\textbf{Model} &\textbf{DFT}& \textbf{MEAM}  & \textbf{MACE} & \textbf{MatterSim} & \textbf{CHGNet} & \textbf{Nequix} & \textbf{SevenNet} &\textbf{ORB} &\textbf{PET}\\
			\textbf{version}&           & \textbf{2018b} &\textbf{MP-0b2} &\textbf{v1.0.0-5M} &\textbf{0.3.0} &\textbf{MP-1} &\textbf{L3I5} &\textbf{v3-C.} &\textbf{MAD}\\
			                &           &                &\textbf{(large)}&                   &               &              &              & \textbf{OMAT} &             \\
			\hline      a   & 8.890	& 8.907 & 9.244 & 9.286 & 9.206 & 9.264 & 9.210 & 9.107 & 8.962\\
						b   & 8.890	& 8.907 & 9.244 & 9.286 & 9.206 & 9.264 & 9.210 & 9.107 & 8.962\\
			$-E_{c}$        & 3.062 &  2.830 & 2.903 & 2.250 & 2.331 & 2.787 & 2.803 & 1.989 & 3.063\\
			$d$             & 2.897 &  2.894 & 2.959 & 2.965 & 2.942 & 2.960 & 2.943 & 2.937 & 2.907\\
			$h$             & 3.404 &  3.617 & 3.277 & 3.274 & 3.270 & 3.261 & 3.241 & 3.348 & 3.273\\
			$\mathrm{C_{11}}$ & 33.0&  16.7  & 29.3  &  30.9 & 29.1  & 30.0  & 28.5  & 40.5  & 80.4\textsuperscript{$\ddag$}\\
			$\mathrm{C_{22}}$ & 33.0&  16.7  & 29.3  &  30.9 & 29.1  & 30.0  & 28.5  & 40.5  & 81.0\textsuperscript{$\ddag$}\\
			$\mathrm{C_{12}}$ & 15.8&   7.3  & 9.5   &  11.7 & 13.1  & 12.0  &  7.7  & 11.3  & 25.1\\
			$\mathrm{C_{33}}$ & 8.6 &   4.7  & 9.9   & 	 9.6 &  8.0  &  9.0  & 10.4  & 14.6  & 27.6\\
			$K_{I}$           &48.8 &  23.9  & 38.8  &  42.6 & 42.2  & 42.0  & 36.2  & 51.8  &105.8\\
			$K_{II}$          &17.3 &   9.4  & 19.8  &  19.2 & 16.0  & 18.0 & 20.8 & 29.1 & 55.6\textsuperscript{$\ddag$}\\
			$K_{III}$         &17.3 &   9.4  & 19.8  &  19.2 & 16.0  & 18.0  & 20.8  & 29.1  & 55.3\textsuperscript{$\ddag$}\\
			MAPE$_\textrm{LHD}$      &       & 29.589& 12.170 & 10.574 & 9.473 & 7.639 & 15.570 & 27.321 & 94.353\\
			{MAPE$^\textrm{S}_\textrm{LHD}$} & &1.678&  3.459 &  3.774 & 3.157 & 3.700 &  3.399 &  1.982 &  1.463\\
			{MAPE$^\textrm{M}_\textrm{LHD}$} &&49.210& 20.411 &	33.854 &30.520 &16.344 & 27.986 & 66.258 &140.844\\
			\hline
			$\sum$MAPE$_\textrm{7Sn}$             &       &264.400& 206.170& 200.270&251.441&148.782&271.699 &276.281 &857.523\\
			{$\sum$MAPE$^\textrm{S}_\textrm{7Sn}$}&       & 19.533&  41.948&  48.049& 92.876& 35.957& 86.296 & 12.919 & 19.157\\
			{$\sum$MAPE$^\textrm{M}_\textrm{7Sn}$}&       &413.036& 295.325& 397.200&436.740&235.242&388.655 &582.493 &1271.379\\
			\hline
		\end{tabular}
	\begin{tablenotes}[para,flushleft]
		\raggedright
		\small
		\textsuperscript{$\ddag$} {Potential does not reproduce the required isotropy of the elasticity tensor ($\mathrm{C_{11}}\neq\mathrm{C_{22}}$}~\&$/$$\wedge$~$K_{II}$$\neq$$K_{III}$). 
	\end{tablenotes}
\end{table}

\begin{table}[!h]
	\caption{Aggregated {total} mean absolute percentage error~({MAPE}$_\textrm{A}$)\,(\%), {aggregated structural mean absolute percentage} 
		error~{MAPE}$^\textrm{S}_\textrm{A}$\,(\%), aggregated mechanical mean absolute percentage error~{MAPE}$^\textrm{M}_\textrm{A}$\,(\%), standard deviation $\sigma$.}
\label{tab:MAPE}
\centering
\setlength{\tabcolsep}{1.0pt} 
	\begin{tabular}{c c c c c c c c c}
		\hline
		\textbf{Model}  &            \textbf{Best}     & \textbf{MACE}  & \textbf{MatterSim} & \textbf{CHGNet} & \textbf{Nequix} & \textbf{SevenNet} &\textbf{ORB} &\textbf{PET}\\
		\textbf{version}&           \textbf{Classical} &\textbf{MP-0b2} &\textbf{v1.0.0-5M} &\textbf{0.3.0} &\textbf{MP-1} &\textbf{L3I5} &\textbf{v3-C.} &\textbf{MAD}\\
		&           \textbf{Potential} &\textbf{(large)}&                   &               &              &              & \textbf{OMAT} &             \\
		\hline
		$\sum${MAPE}$_\textrm{A}$              & 153.398 & 128.163 & 132.980 & 210.580 & 117.269 & 169.717 & 171.400 & 401.684\\
		$\sigma$                                           & 34.153  &  25.543 &  23.766 &  24.084 &  17.212 &  33.412 &  34.480 & 124.583\\		
		{$\sum${MAPE}$^\textrm{S}_\textrm{A}$} &  20.415 &  28.364 &  26.706 &  56.316 &  31.806 &  25.687 &  11.024 &  20.911\\
		$\sigma$                                           &   2.294 &  5.782  &   6.569 &  12.426 &   4.709 &   5.072 &   1.655 &   3.463\\
		{$\sum${MAPE}$^\textrm{M}_\textrm{A}$} & 269.989 & 203.545 & 278.844 & 358.471 & 204.951 & 275.318 & 381.365 & 608.667\\
		$\sigma$                                           &  49.080 &  33.975 &  45.368 &  43.640 &  26.407 &  46.471 &  70.089 & 182.754\\
		\hline
	\end{tabular}
\begin{tablenotes}[para]
	\small
\end{tablenotes}
\end{table}

\begin{table}[!h]
	\caption{Relative performance measured as normalized seconds/timestep in molecular dynamics (MD) simulation.}
	\label{tab:time}
	\centering
	\setlength{\tabcolsep}{1.0pt} 
	\tiny
		\begin{tabular}{c c c c c c c c c c c c}
			\hline \multicolumn{12}{c}{MoS$_2$} \\ 
			\hline 
			\textbf{Model} &\textbf{SW}   & \textbf{REBO} & \textbf{MACE} & \textbf{MACE} & \textbf{MACE} & \textbf{MatterSim} & \textbf{CHGNet} & \textbf{Nequix} & \textbf{SevenNet} &\textbf{ORB} &\textbf{PET}\\
			\textbf{version}&\textbf{2016}&               &\textbf{MP-0b2} &\textbf{MPA-0} &\textbf{MP-0b2} &\textbf{v1.0.0-5M} &\textbf{0.3.0} &\textbf{MP-1} &\textbf{L3I5} &\textbf{v3-C.} &\textbf{MAD}\\
							&             &               &\textbf{(small)}&\textbf{(medium)} &\textbf{(large)}&                   &               &              &              & \textbf{OMAT} &             \\
			\hdashline  
			s/timestep & 1 & 13 & 19526 & 52596 & 120777 & 2495 & 4924 & 3090 & 8838 & 29917 & 5603\\
			
			\hline \multicolumn{12}{c}{Si} \\ 
			\hline
			\textbf{Model} &\textbf{SW}   & \textbf{MEAM} & \textbf{MACE} & \textbf{MACE} & \textbf{MACE} & \textbf{MatterSim} & \textbf{CHGNet} & \textbf{Nequix} & \textbf{SevenNet} &\textbf{ORB} &\textbf{PET}\\
			\textbf{version}&\textbf{1985}& \textbf{2011} &\textbf{MP-0b2} &\textbf{MPA-0} &\textbf{MP-0b2} &\textbf{v1.0.0-5M} &\textbf{0.3.0} &\textbf{MP-1} &\textbf{L3I5} &\textbf{v3-C.} &\textbf{MAD}\\
							&             &               &\textbf{(small)}&\textbf{(medium)} &\textbf{(large)}&              &               &              &              & \textbf{OMAT} &             \\
			\hdashline
			s/timestep & 1 & 2 & 10000 & 28000 & 65000& 990 & 1463 & 1578 & 3367 & 11447 & 2562 \\
			\hline
		\end{tabular}
\end{table}
\clearpage

\bibliography{MLIPs}

\end{document}